\documentclass[preprint2]{aastex}

\usepackage{natbib}
\usepackage{color}

\slugcomment{Not to appear in Nonlearned J., 45.}

\shorttitle{Sausage-pinch Instability}
\shortauthors{Srivastava et al.}

\newcommand{\be}{\begin{equation}}
\newcommand{\ee}{\end{equation}}

\begin{document}


\title{Observational Evidence of Sausage-Pinch Instability in Solar Corona by SDO/AIA}


\author{A.K.~Srivastava\altaffilmark{1}}
\affil{$^1$Aryabhatta Research Institute of Observational Sciences (ARIES), Manora Peak, Nainital-263 129, India}

\author{R.~Erd\'elyi\altaffilmark{2}}
\affil{$^2$Solar Physics and Space Plasma Research Centre (SP2RC), School of Mathematics and Statistics, The University of Sheffield, Sheffield, U.K.}

\author{Durgesh~Tripathi\altaffilmark{3}}
\affil{$^3$Inter-University Centre for Astronomy and Astrophysics, Post Bag 4, Ganeshkhind, Pune 411007, India.}

\author{V.~Fedun\altaffilmark{2,4}}
\affil{$^2$Solar Physics and Space Plasma Research Centre (SP2RC), School of Mathematics and Statistics, The University of Sheffield, Sheffield, U.K.}
\affil{$^4$Department of Automatic Control and Systems Engineering, The University of Sheffield, Mappin Street, Sheffield S1 3JD, U.K.}

\author{N.C.~Joshi\altaffilmark{1}, P.~Kayshap\altaffilmark{1}}
\affil{Aryabhatta Research Institute of Observational Sciences (ARIES), Manora Peak, Nainital-263 129, India.}



\begin{abstract}
We present the first observational evidence of the evolution of sausage-pinch instability in Active Region~11295 during a prominence eruption using data recorded on 12 September 2011 by the Atmospheric Imaging Assembly (AIA) onboard the Solar Dynamics Observatory (SDO). We have identified a magnetic flux tube visible in AIA 304 \AA\ that shows curvatures on its surface with variable cross-sections as well as enhanced brightness. 
These curvatures evolved and thereafter smoothed out within a time-scale of a minute. 
The curved locations on the flux tube exhibit a radial outward enhancement of the surface of about 1-2 Mm (factor of 2 larger than the original thickness of the flux tube) 
from the equilibrium position. AIA~193~{\AA} snapshots also show the formation of bright knots and narrow regions inbetween at the four locations as that of 304 \AA\ along the flux tube where plasma emission is larger compared to the background. 
The formation of bright knots over an entire flux tube as well as the narrow regions in $<$ 60 s may be the morphological 
signature of the sausage instability. 
We also find the flows of the confined plasma in these bright knots along the field lines, which indicates the dynamicity of the flux tube that 
probably causes the dominance of the longitudinal field component over short temporal scales. 
The observed longitudinal motion of the plasma 
frozen in the magnetic field lines further vanishes the formed curvatures 
and plasma confinements as well as growth of instability to stablize the flux tube.
\end{abstract}



\section{Introduction}
A wide range of MHD instabilities have been observed in 
solar atmosphere in recent years in association with various dynamical processes
\citep[e.g.][and references therein]{Williams2005, Srivastava2010, Foullon2011}. 
\cite{Kumar2010} discovered evidence of  coalescence instability for loop-loop interaction in association with M7.9/1N class solar flare 
as previously theoretically investigated in great details \citep[e.g.][]{Tajima1982, SakaideJager1996}.
Kliem et al. (2000) have also reported the dynamical reconnection scenario by repeated 
formation and subsequent coalescence of magnetic islands in solar active regions.
Recently, 
\cite{Innes2012} interpreted the break-up and finger like structures in down-flowing plasma following 
a filament eruption on 7 June 2011 as evidence of Rayleigh-Taylor instability. 
\cite{Hillier12a} and \cite{Hillier12b} have numerically modelled Rayleigh-Taylor instability
in the solar prominences to explain the observed plasma upflows and reconnection
triggered downflows.
Ballooning and torus instabilities have also 
been reported as one of the prominent 
mechanisms for flux rope eruptions 
\citep{Aulanier2010}. 
\citet{Tsap2008} have reported the generation of ballooning instability due to the 
confined kink instability in ideal MHD regime applicable to the solar atmosphere as previously 
analytically modelled by, e.g., \cite{Hood79}.
Furthermore, Kelvin-Helmholtz and Alfv\'en instabilities in 
the solar corona are also observed and theorized 
by various authors \citep{Foullon2011, OfmanThompson2011, Taroyan2011}.  
Motivated by the observations of these instabilities using here excellent high-resolution data from the 
Transition Region and Coronal Explorer (TRACE), Hinode and SDO missions, a considerable effort has gone into analytical and numerical 
modelling in order to study the characteristics of the different types of instabilities \citep[see e.g.]
[and references cited therein]{Torok2004,TorokKliem2005,Haynes2007,Taroyan2011,Soler2010,Zaqarashvili2010,Botha2012}.


The kink-unstable mode ({\it m}=1) of cylindrical magnetic flux tubes may seem to 
be observed frequently in the solar corona \citep{Srivastava2010}. However, 
another type of instability known as sausage instability ({\it m}=0) mode, 
which is theoretically investigated in astrophysical plasma
\citep{Priest1982, Sturrock1994, Aschwanden2004}, 
to the best of our knowledge, has not yet been observed 
in solar atmosphere. 
In a cylindrical magnetic flux tube, the inward directed 
Lorentz force is counteracted by increasing 
the gas plasma pressure 
gradient that is directed outward, and evolution of such 
instability takes place after meeting certain criteria, i.e., 
${B_{\theta}}/{B_{z}}$$>$$1.4$ \citep[see][]{Kadomtsev1966, Aschwanden2004}.
In the present paper, we outline the observational evidence of sausage pinch instability 
evolving in an eruptive 
flux tube from the Active Region NOAA AR~11295 on 12 September 2011.
 In section 2 we 
describe the observations. In section 3 we report the detection of sausage-pinch 
instability in corona. The last section contains the summary and discussions.

\section{Observations}
A C9.9 class solar flare was observed during the emergence of active region AR11295 (N21E59) at the 
North-East limb on 12 September 2011. The flare started at 20:30 UT with peak intensity at 20:54 UT, and
ended at 21:51 UT.
We have used 
imaging data showing the dynamics of AR~11295 as observed by the SDO/AIA, 
which has a maximum resolution of 
0.6$"$ per pixel and a cadence of 12 s. AIA provides full disk 
observations of the Sun in three ultra-violet (UV) 
continua at 1600 \AA, 1700 \AA, 4500 \AA, and seven Extreme 
Ultra-Violet (EUV) narrow bands at 171 \AA, 
193 \AA, 211 \AA, 94 \AA, 304 \AA, 335 \AA, and 131 \AA\, 
respectively (Lemen et al. 2012). Therefore, it provides observations 
of multi-temperature, high spatial and temporal resolution  plasma dynamics of 
solar atmosphere. 
Here, we use the data recorded 
by two filters of AIA, namely, 304 \AA\  and 193 \AA . The images recoded in 
304~{\AA}~(max. formation temperature T$_{f}$=10$^{5}$ K) provides information about the plasma 
dynamics in upper chromosphere and lower transition region, whereas those in 
193~{\AA} (T$_{f}$=1.58$\times$10$^{6}$ K) reveal information about the lower corona. The 
primary contributing ion and the temperature response for each of these channels can vary, depending on the 
plasma features on the Sun's surface being observed 
 \citep{ODwyer2011,Del11}. 
The time-series of SDO/AIA 
data has been reduced by the SSW cutout service \footnote{http://lmsal.com/get\_aia\_data/}.

\section{Detection of Sausage-pinch Instability}

\begin{figure*}
\centering
\mbox{
\includegraphics[scale=0.53, angle=90]{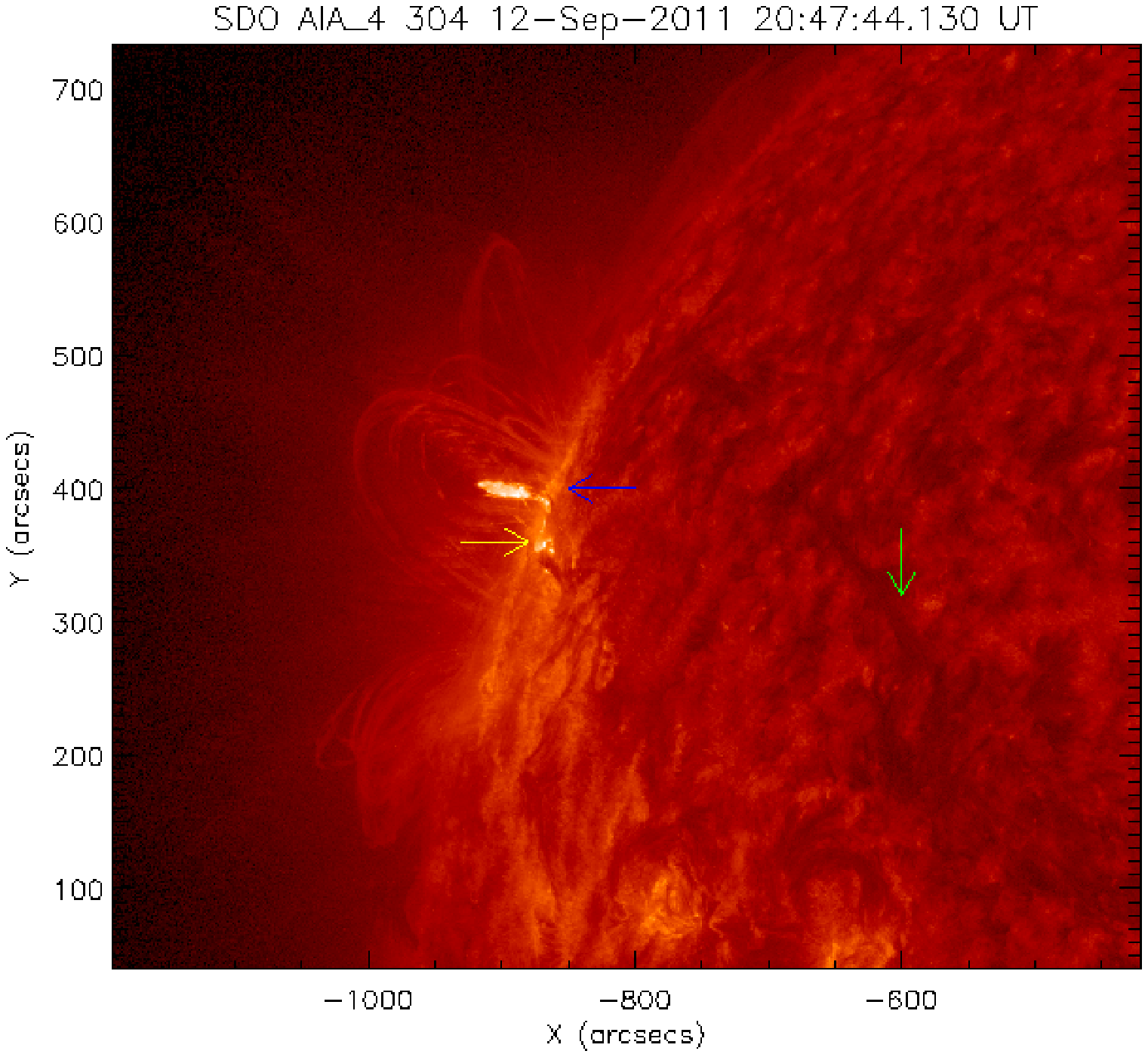}
\includegraphics[scale=0.53, angle=90]{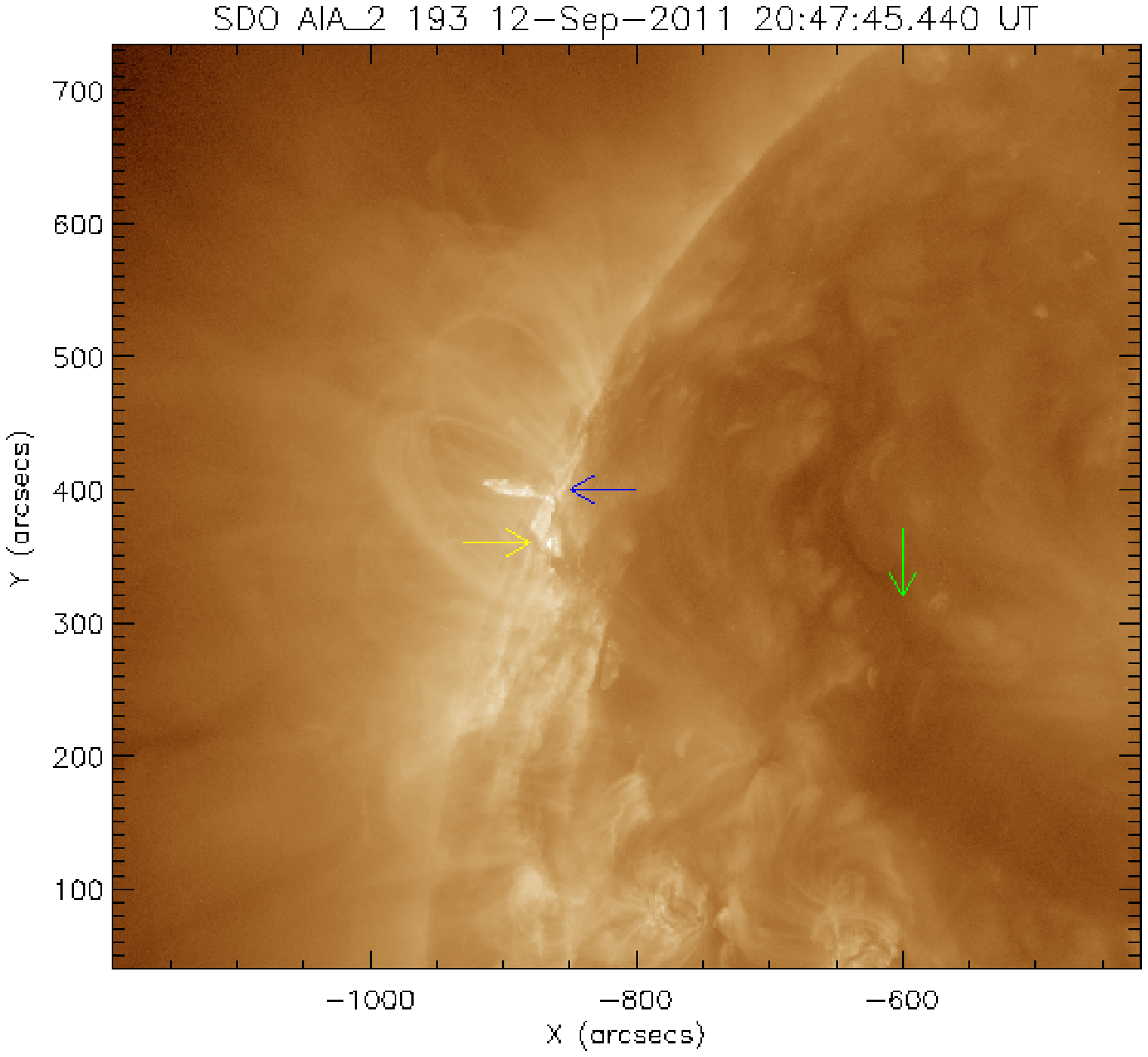}}
\caption{\small
SDO/AIA 193 \AA\ (right) and 304 \AA\ (left) EUV images showing the
filament channel (green-arrow), activated filamentary part
(yellow-arrow), rising instable flux-rope and 
overlying loops (blue-arrow) in AR11295 well before the flare.  
}
\label{fig:JET-PULSE_1}
\end{figure*}

Fig.~~\ref{fig:JET-PULSE_1} shows the 
images of the active region AR~11295 and the 
regions in their vicinity at the North-East limb recorded 
using 304~{\AA} (left panel) and 193~{\AA} 
filters (right panel). Two filament channels are visible in these snapshots 
highlighted by green and yellow arrows. 
The eastward filament channel marked by yellow arrow becomes partially activated around 20:45 UT, 
well before the start of the flare. The filament passes through different topological changes and shows 
plasma heating 
before it finally kinks and erupts partially
(cf., MFT.mpeg). 
This eruption is classified as partial eruption 
\citep[e.g.][]{Gil00, Gibson2006, Tri07, Tripathi2009, Tripathi2013}. 

\begin{figure*}
\centering
\mbox{
\includegraphics[scale=0.35, angle=90]{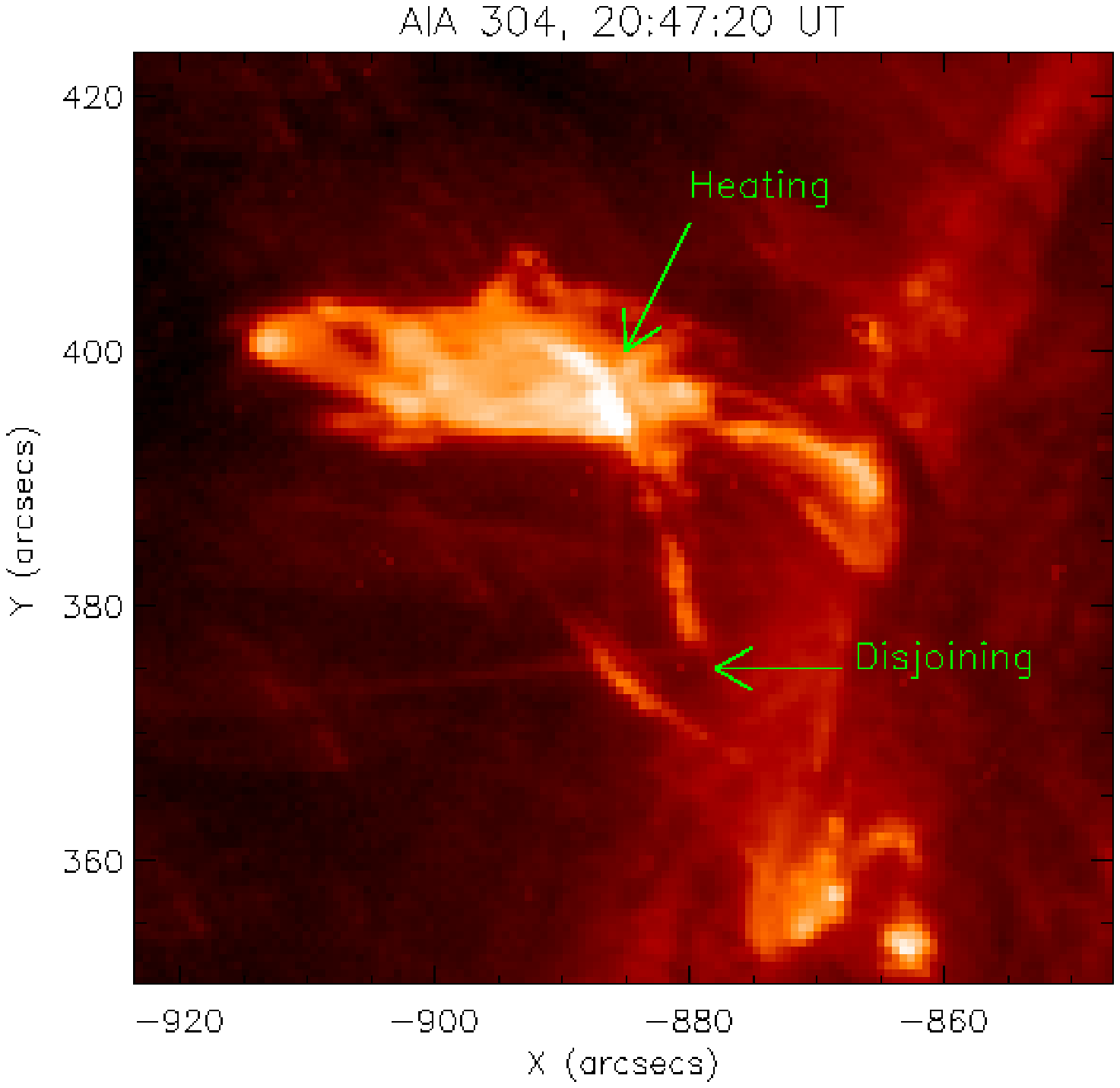}
\includegraphics[scale=0.35, angle=90]{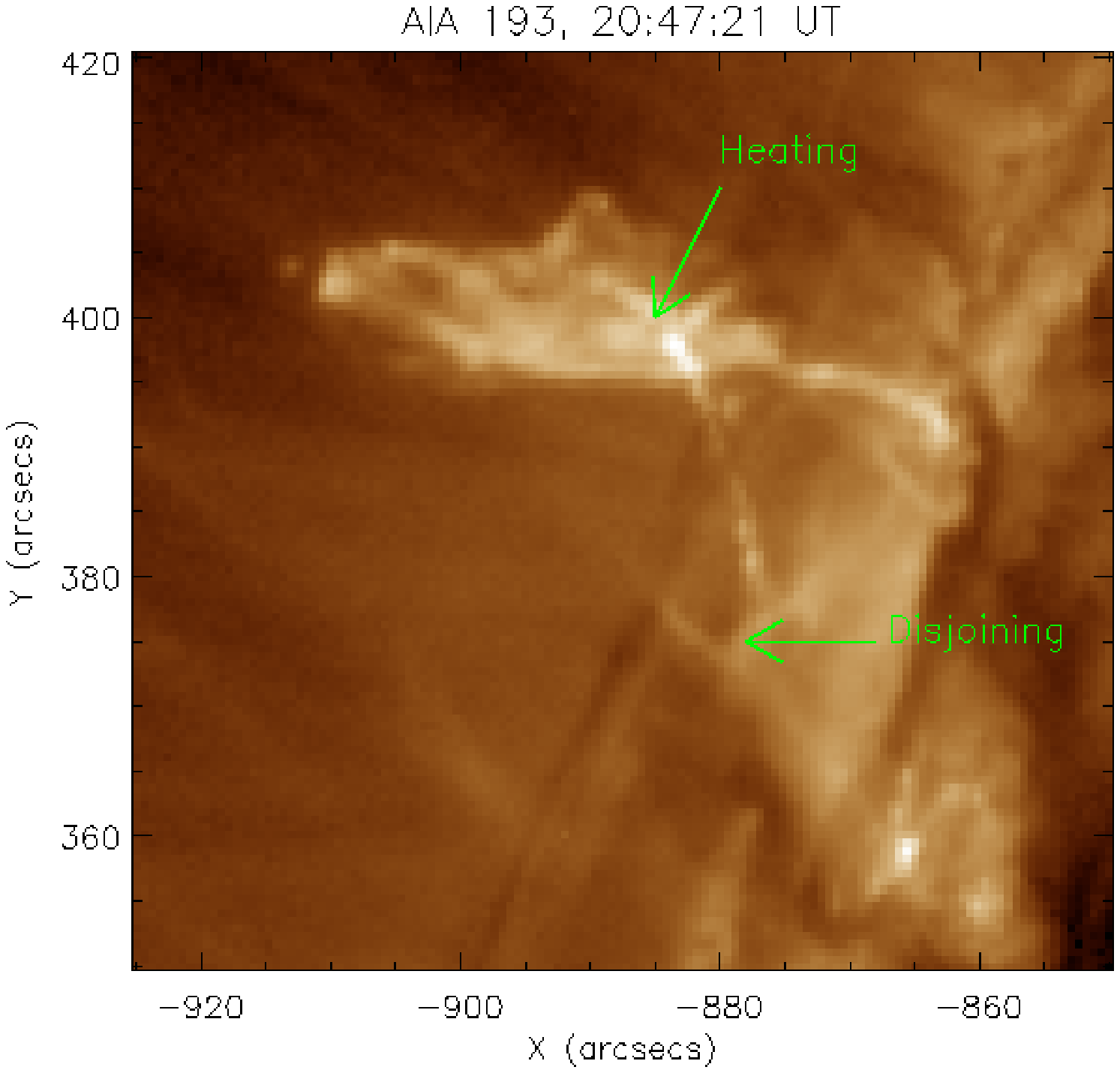}
}
\mbox{
\includegraphics[scale=0.35, angle=90]{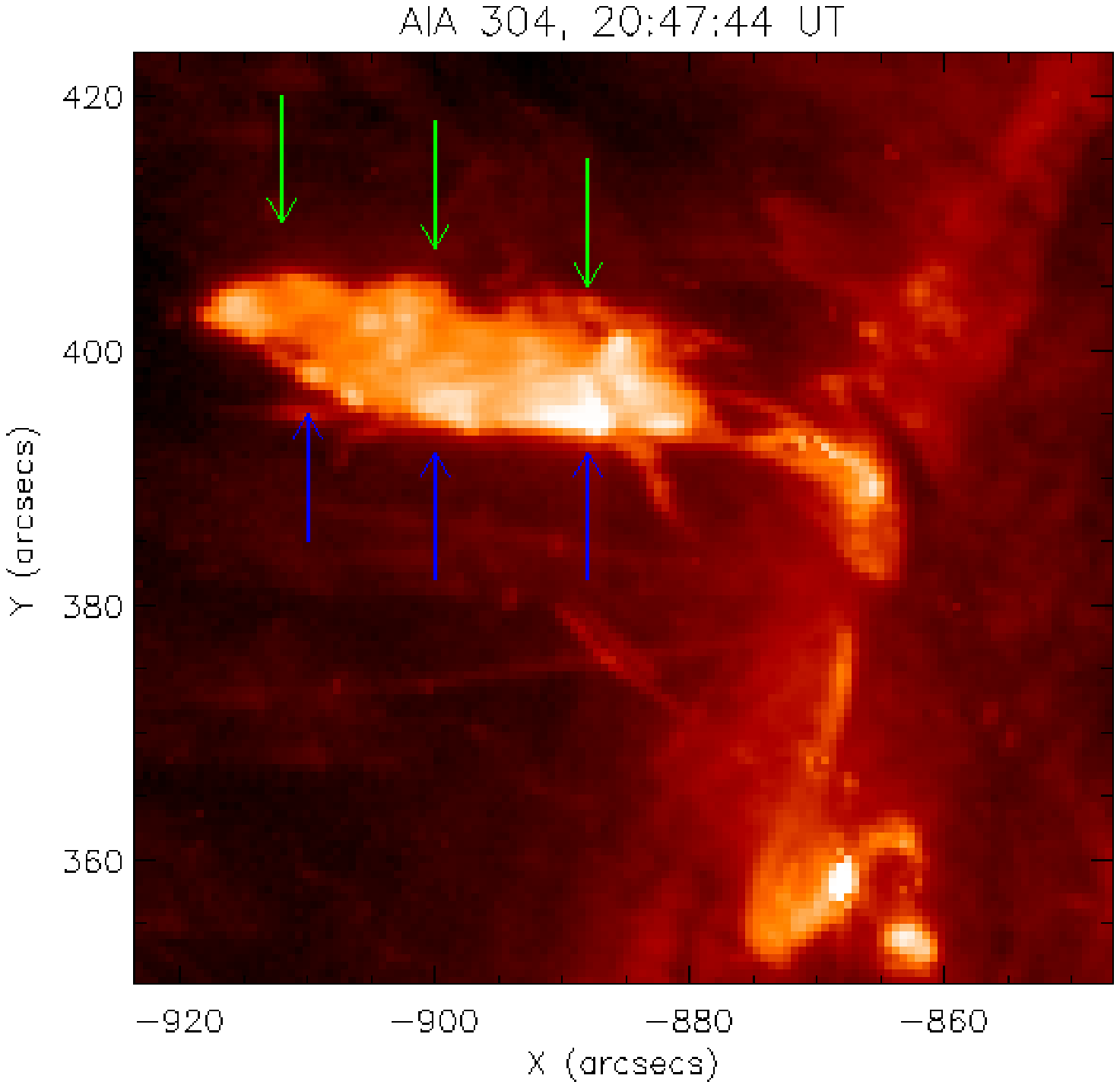}
\includegraphics[scale=0.35, angle=90]{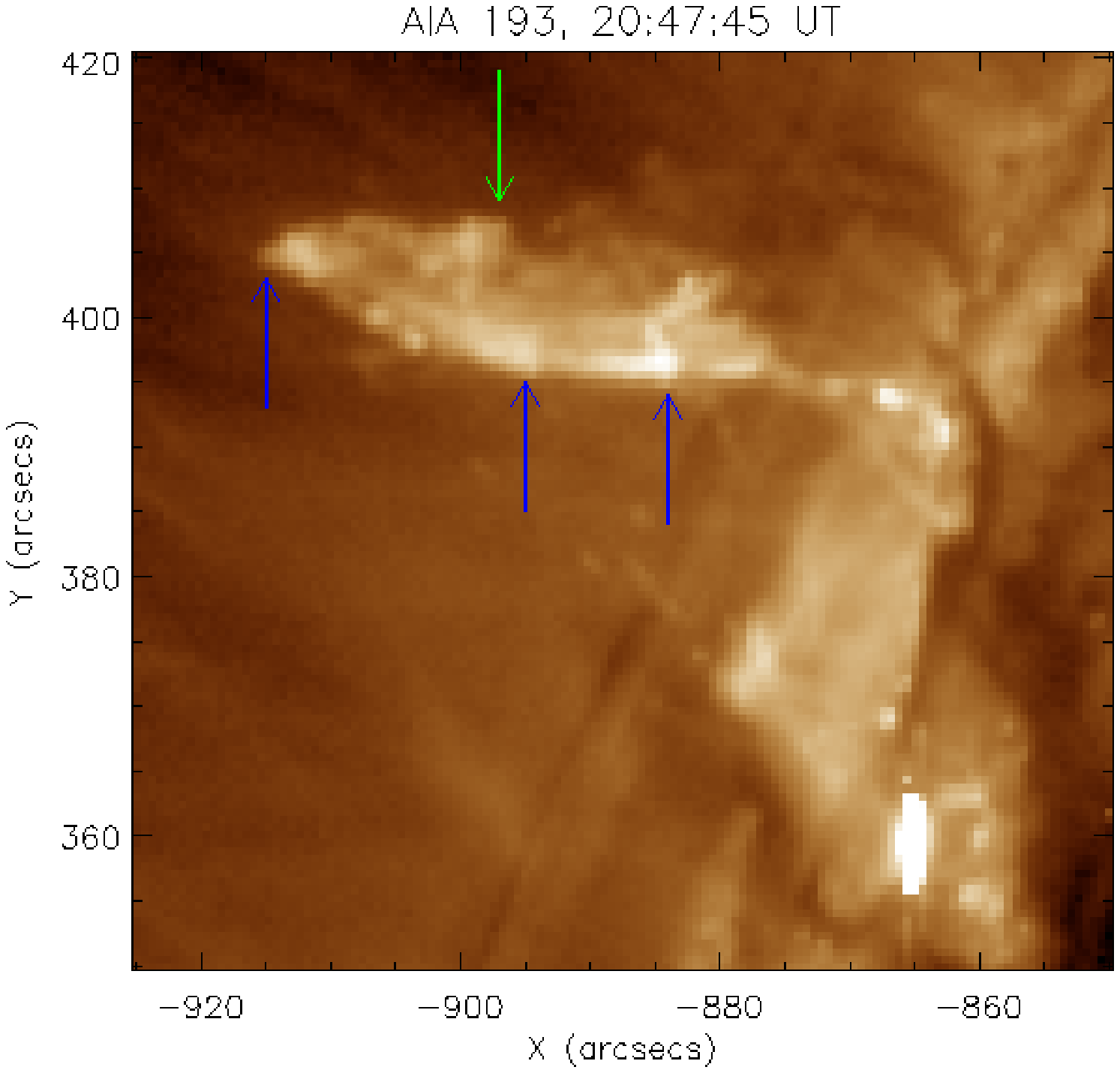}
}
\mbox{
\hspace{0.8cm}
\includegraphics[scale=0.35, angle=90]{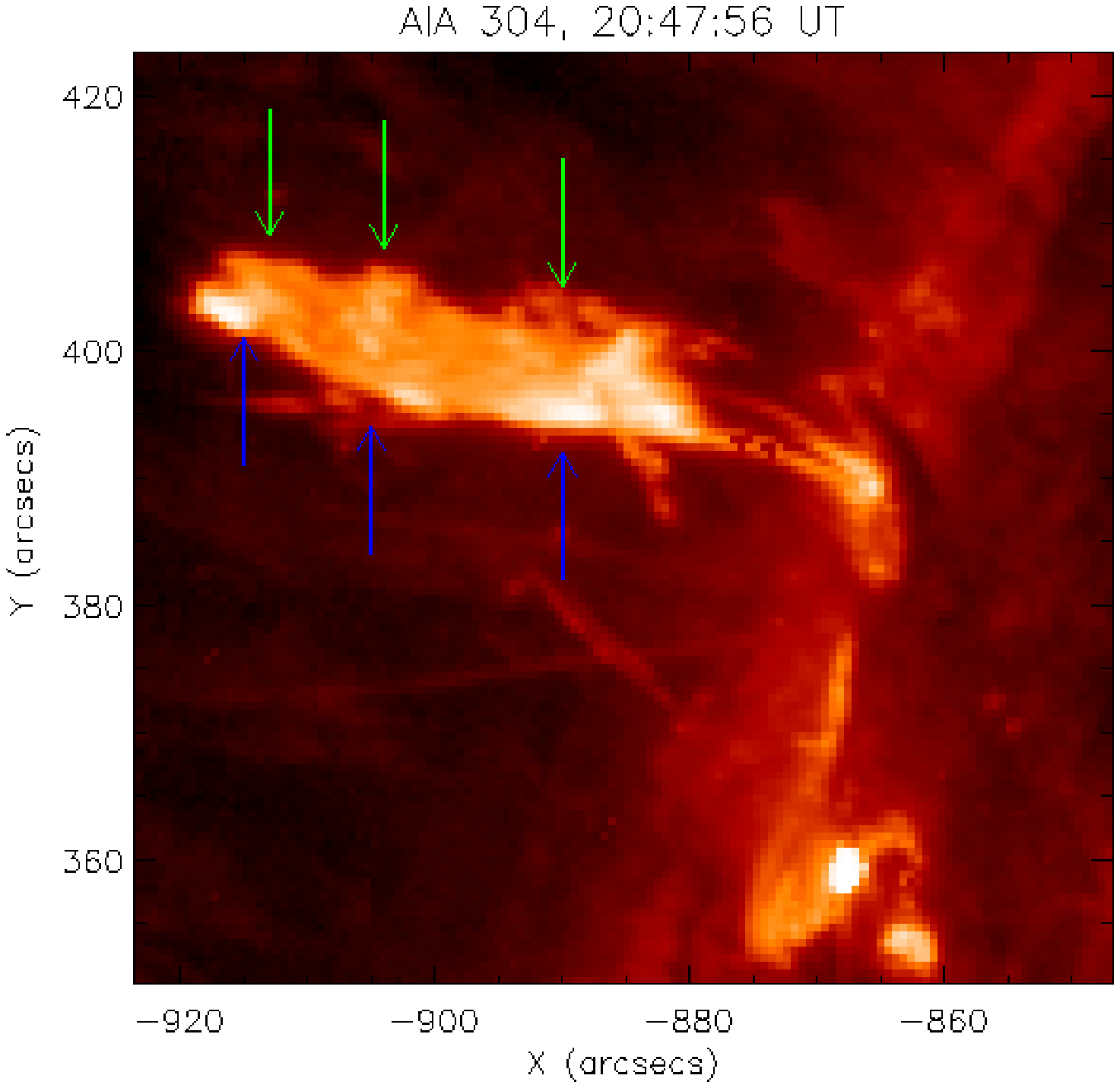}
\hspace{-0.02cm}
\includegraphics[scale=0.35, angle=90]{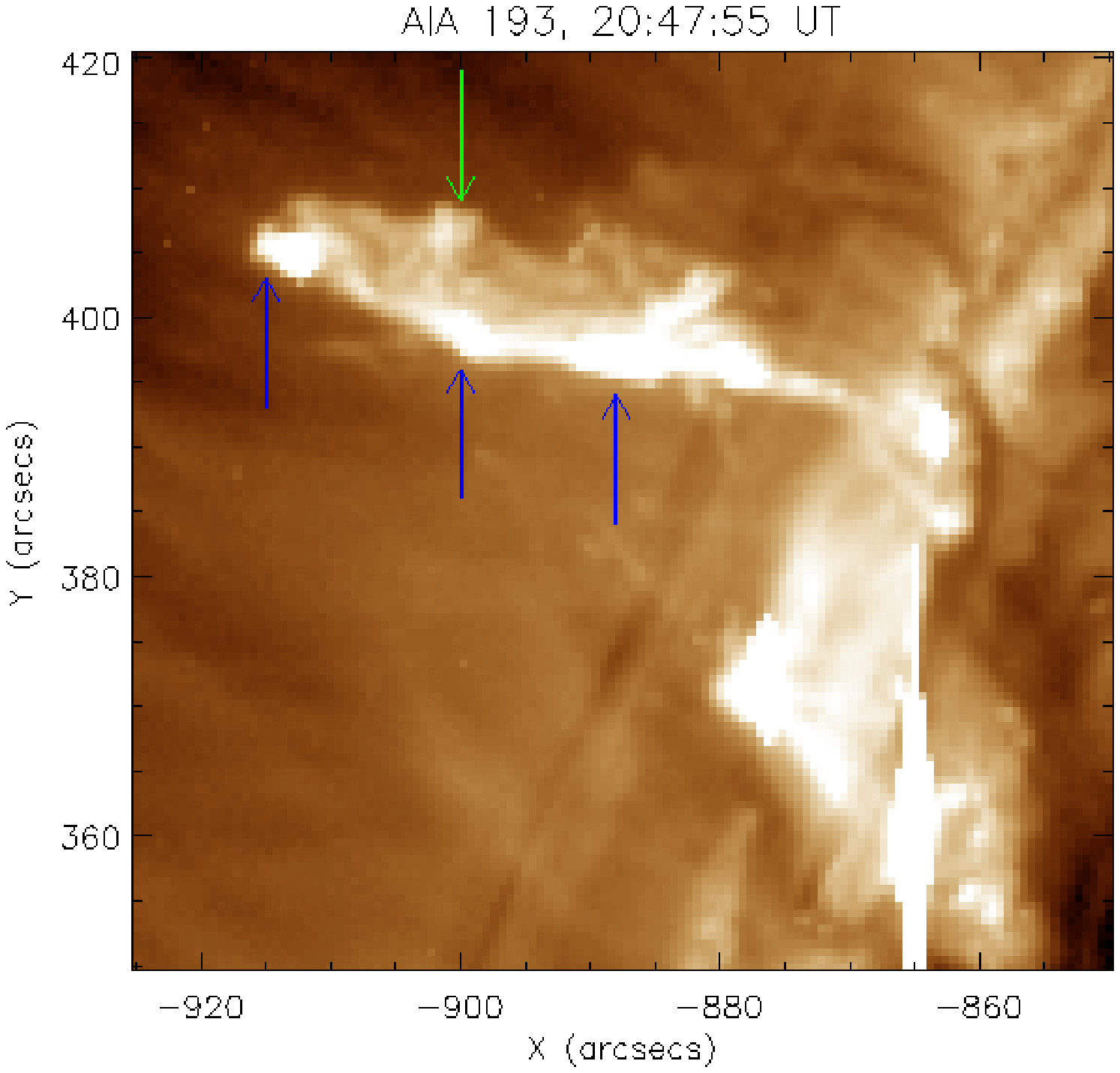}
\thicklines 
 $\color{green} \put(-76,82){A}$ 
 $\color{green} \put(-112,90){B} $ 
 $\color{green} \put(-145,100){C} $ 
 $\color{blue} \put(-106,120){D} $ 
\thicklines 
 $\color{green} \put(-260,75){A}$ 
 $\color{green} \put(-286,80){B} $ 
 $\color{green} \put(-324,93){C} $ 
 $\color{blue} \put(-294,113){D} $ 
}
\mbox{
\includegraphics[scale=0.35, angle=90]{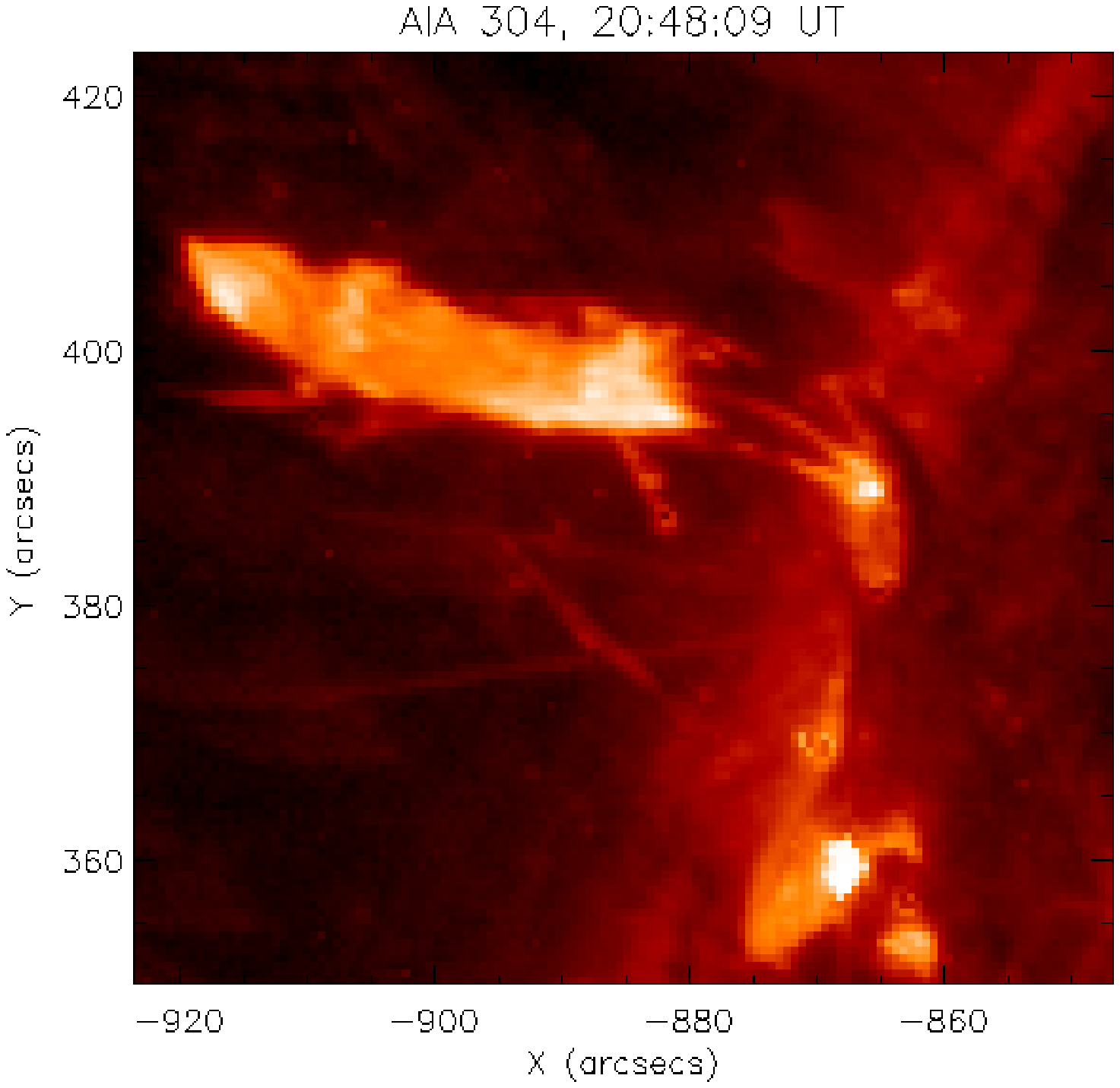}
\includegraphics[scale=0.35, angle=90]{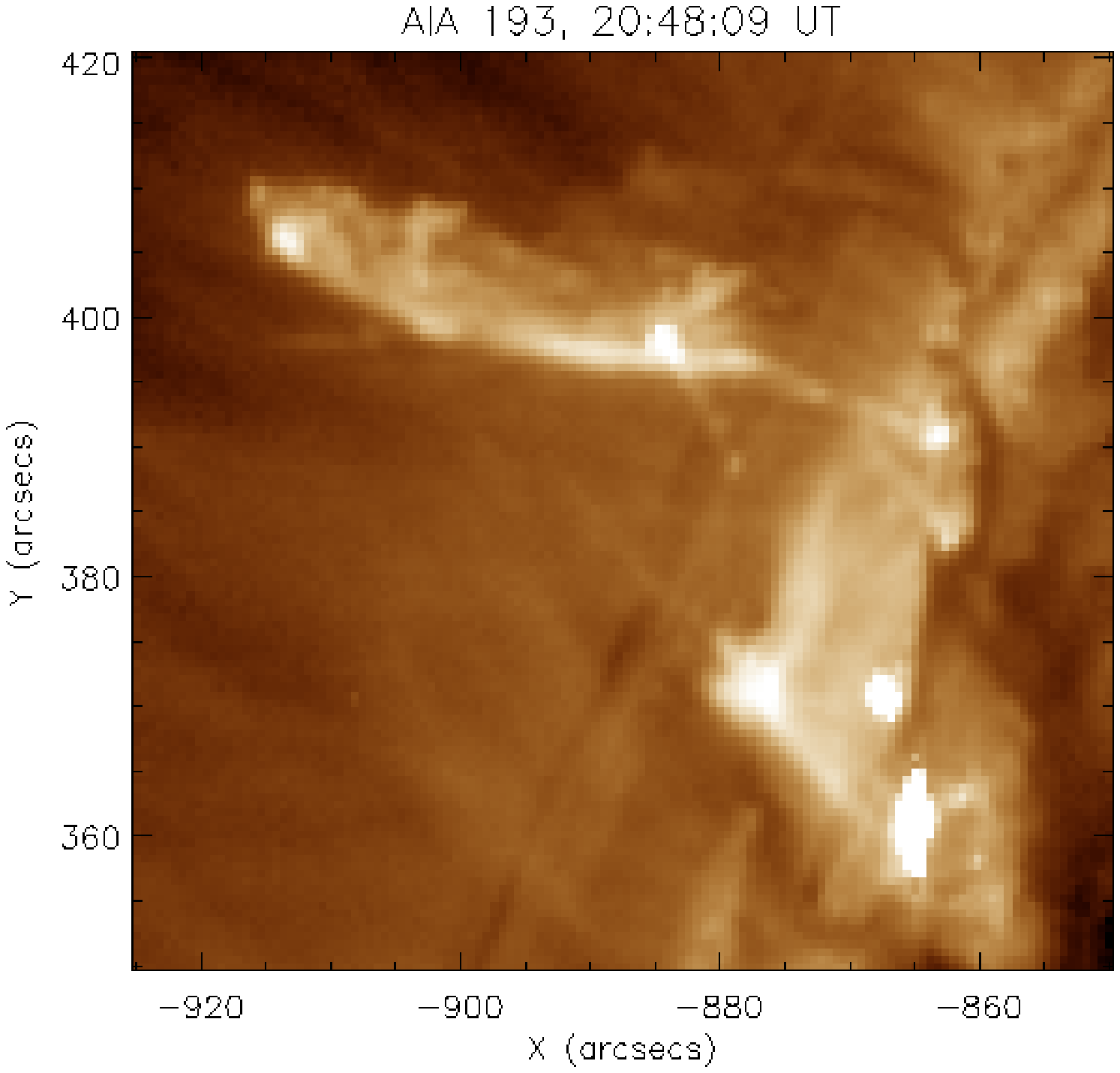}
}
\caption{\small SDO/AIA 304 \AA\ and 193 \AA\ temporal image data showing the 
evolution of sausage-pinch instability. The formation of enhanced density as well as
increased cross-sections are clearly evident over the fluxtube, which is overlying the 
complex active region. The three blue arrows on 193 \AA\ (20:47:55 UT) and 304 \AA\ (20:47:57 UT) snapshots respectively 
mark (right to left) the positions 'A', 'B', and 'C', while the green arrow marks the position
'D', where surface curvatures are evolved and rippled out. %
}
\label{fig:JET-PULSE_2}
\end{figure*}

In Fig.~\ref{fig:JET-PULSE_2}, the left and right 
vertical columns, respectively, show the time-series of 
the partial and zoomed FOV of a rising flux tube 
(part of the erupting prominence) in AR~11295 
for a rather short duration of 20:47-20:48 UT in 304 
\AA\ and 193 \AA. During this period, we note that, 
the flux tube does not rise much as is evident from 
Fig.~\ref{fig:JET-PULSE_2} despite there are considerable 
topological changes as well as plasma flows. The top row in Fig.~\ref{fig:JET-PULSE_2} 
demonstrates the 
internal reconnection above the two footpoints of the flux tube and energy release
via kinking. The southern foot-point of the flux tube becomes 
disconnected during the evolutionary process
due to 
reconnection, and its 
connectivity is changed. This facilitates the 
slow rise of the flux tube in the corona. 

The twist is transferred in the remaining part of the magneto-plasma 
of the flux tube. It should also be noted that the enveloping closed 
flux tube is overlying the complex magnetic field 
configuration and associated plasma in the core of the active region. 
The flux tube reveals concave and non-static curvatures at various locations on surface 
(cf., 304 \AA\ snapshot in the second row) on both side. 
 We have marked these locations by green (northern side) and 
blue (southern side) arrows. The concave curvatures appear like bright 
knots due to the enhance emission at those locii. 
This could be explained in terms of line-of-sight (LOS) integration of the emission. 
It is also observed that the narrow regions also become 
brighter along with knots. This may be due to the highly 
dynamic nature of flux tube and its plasma that is flowing
along with field lines (cf., Fig.~5).
Apart from sub-mega Kelvin (304 \AA~) and mega-Kelvin (193 \AA~)
AIA channels, we also examined a more hotter flare channel of 131 \AA~
in which co-temporal data is available. The appearance of the same enhanced
areas, as visible in 304 \AA~in Fig.~2, are also evident in this particular band in the enveloping flux tube.
The flare energy is enhanced in the low-lying southward
directed loop systems (not shown here) in the same active region, although, it 
is not directly related with the observed flux tube.
The plasma may be multi-thermal leading us to see that in the 
multi-channels of AIA. However, the extent to which this emission 
is seen in multiple co-temporal channels is almost the same. Therefore, it is plausible to
conclude that these bright knots are not due to heating effects.

At the locii where the cross-section of the flux tube has increased 
more plasma may be accommodated. 
While, other locii where the flux tube is 'pinched' and has smaller 
cross-sections may have rarefaction and less plasma density 
\citep{Syr81}. Since the radiation in 
the corona is essentially optically thin, the locii with enhanced 
cross-sections appear brighter. This phenomenon can also be seen in 
the 193 \AA\ (second row, right panel) image and is shown by the arrows. 
From the images in the third row, this is further evident in both 
wavelengths 
that the increased cross-sections 
in form of knots are more evolved 
(cf., 20:47:55 UT snapshots). The phenomenon is well observed 
in some particular AIA channels in which the LOS contribution 
of the plasma is significant in filling these regions of the 
flux tube. Moreover, this is very short duration dynamics observed
by few AIA channels successfully and co-temporally. We note here that these 
curvatures faded away within $\sim$1.0 minutes of their evolution 
from the entire body of the flux tube. The plasma 
necks (refer to narrow regions in the current resolution 
of AIA) are formed in-between these knots, which are the 
pinched regions.
However, the exact quantification of flux tube necks, i.e., decrease in loop width 
opposite to the sausage knots, is not plausible in the observed thin and enveloping flux tube of radius 1 Mm 
considering the current resolution of AIA (0.6$"$/pixel).  

We interpret these observations as the most likely evolution of  
sausage-pinch instability in the early phase of partially eruptive flux tube in NOAA AR11295. 
The flux tube becomes unstable at the 
knot locations (cf., schematics overlaid on AIA snapshots in Fig.~3)
where the confining field is most likely concave and 
the longitudinal field is less dominant over the local azimuthal component. 
This leads to an instability known as the sausage pinch instability. 
The Lorentz force, and therefore the pinch-instability in a plasma cylinder, 
could be generated either by an axial current or by presence of an ambient 
longitudinal magnetic field. In the first situation, in the plane which 
is perpendicular to the axis, the internal current $j_z$ generates an 
azimuthal magnetic field $B_\theta$ around itself. Interaction of the 
axial current with the induced magnetic field leads to excitation of 
the {\it z}-pinch instability of the plasma cylinder (cf., Fig.~3, top-left panel). 
On the other hand, the longitudinal magnetic field 
inside the flux tube when varies along its radius at a particular
height, then induces an azimuthal current ($j_\theta$).
During interaction of this current with the ambient magnetic field, an inwardly 
directed Lorentz force is generated that confine the plasma in cylindrical 
geometry. This effect is known as $\theta$-pinch instability (cf., Fig.~3, top-right panel). 
At the locii where the plasma column is pinched it can be stabilized against 
the sausage {\it z}-pinch  instability by a longitudinal field that is significantly 
dominant over the azimuthal component \citep[for more details see][]{Kadomtsev1966, Aschwanden2004}. 
The $\theta$-pinch is neutrally stable and, therefore, stability also depends
on additional ambient fields (see e.g. Freidberg 1982).
Either mechanism may enforce the sausage instability at a particular moment 
in the flux tube. However, it should be noted that the present observational base-line can not distinguish between these two mechanisms
and it should be considered as general interpretation of the instability.

The Time-Distance maps (Fig.~4) also reveal the
dynamics of the increased cross-sectional areas over the fluxtube between 
the narrow pinched and straight regions. 
For example, the time-distance diagrams in AIA 304 \AA\ along the slits are drawn across the various
surface curvatures over the flux tube marked by position 'A', 'B' on its southward part
, and 'D' on northward part as shown in SDO/AIA 193 and 304 \AA\ snapshots at 20:47 UT (cf., Fig.~2).
The initial positions of the vertical slits (top-right corners in each Time-Distance diagram) 
are taken at the central core of the active region to which the fluxtube is enveloping. This point 
is considered as reference point, and all distances along the slit are measured w.r.t. this position. 
In time, 
the variation in the surface curvatures (ripples) and their evolution are clearly shown in Fig.~4.
The top{\bf-left} panel of Fig.~4 shows the time-distance plot along the slit placed across curvature 'A'
(cf., Fig.~2).
The surface of the flux tube rises vertically 
up to $\sim$1.0 Mm outward and then, again, subsided back. 
Similarly,
the curvatures at positions 'B', and 'D'  (cf., top-right and bottom-left panels in Fig.~4), 
respectively, show the outward displacements of $\sim$1.5 Mm, $\sim$2 Mm,
in the projection that subsided 
within 1.0 min.  
We exclude the measurement on the knot marked as 'C' in Fig.~2 as this portion 
outgrows obliquely and we may not have true variations in its outer surface curvature.
We notice qualitative signature of the formation of bright knot 
at this position.
The last panel of Fig.~4 shows almost no variation of the surface of clearly evolved neck between knots B and C. 
This region is not associated with any sliding motion of plasma blobs, e.g., as observed between A \& B
and C \& D. A small pinching is also evident that recovers
in time in dynamic flux tube. In this  dynamic flux tube, it is difficult to constrain  purely static pinched regions.
However, this measurement 
shows that such regions may be created for some definite duration. Therefore, non-ideally though under the regime 
of dynamic flux tube, the morphological evidence of sausage-pinch instability incurred.

Kruskal-Shafranov condition for a tube of half length 'L' 
and half-width 'a' that subjects to the internal kink instability
(m=1) mode is

\be
2L>\lambda_{s}=\frac{2\pi a B_{z}}{B_{\theta}}.
\ee

Using the loop morphological parameters (L$\sim$36 Mm, a$\sim$1 Mm) and equation,
we get ${B_{\theta}}/{B_{z}}$$>$$2\pi\,a/2L$$\approx$0.08.
This means that if the conditions for sausage-pinch instability (m=0)
is met (i.e., ${B_{\theta}}/{B_{z}}$$>$$1.4$), the tube is already subjected to the 
kink instability. The bottom-left panel of Fig.~3 also shows the formation 
of kinked flux tube and internal magnetic reconnection. The initial kink instability
later most likely triggers the formation of sausage knots as well as counteracted 
pinched regions. The possible physical mechanisms underlying of these observations are illustrated in Fig.~3.
 
\section{Results and Discussion}\label{SECT:DISS}
In this paper we present observational evidences of the evolution of bright knots and narrow region in between
, as well as  cross-sectional variation at {\bf these} places over a flux tube in AR11295. These variations are seen on a rather short time scale 
of 1.0 min. The radial outward displacements of the concave surfaces of the 
observed flux tube are found to lie within the range of 1-2 Mm. 
Our observations show that the places with 
increased cross-sections are almost double at a time of the normal width of flux tube. 
This is a morphological evidence for the evolution of sausage-pinch instability.
The activation of sausage-pinch instability attempts to disrupt the confined plasma of 
the flux tube in lateral direction, while the pinched regions work oppositely and 
force the plasma along field lines in the longitudinal direction. 

We conjecture that  pinched regions of the observed flux tube in-between the sausage-unstable 
parts are evident as more field-aligned and straighter where the longitudinal 
magnetic field ({\it B$_{z}$}) seems to be dominant over the azimuthal component 
({\it B$_{\theta}$}). These are the 'neck' parts of flux tube in-between the 
'bulky blobs' of sausage-unstable parts (cf., 193 \AA\ snapshot on 20:47:56 UT), 
which is the typical morphological scenario of a sausage-pinch unstable flux tube. 
This is suggestive of the fact that the longitudinal component of the magnetic 
field in the pinched regions becomes dominant.

Since the flux tube is changing 
its geometry and connectivity during eruption, therefore, 
the local active region dynamics and 
magnetoplasma configuration may not allow the full growth of sausage-pinch 
instability in the flux tube
that depends highly on its morphology as well as 
plasma properties \citep{Aschwanden2004}. 
The observed thin flux tube initially seems to be enveloping the 
various other dynamic flux tubes lying below and above of it.
It should be noted that it is not a static flux tube where plasma confinement 
may occur in the form of static sausage and pinched regions.
It was the part of the bulk flux-rope system, which was very complex and dynamic, and containing many
flux tubes interacting and reconnecting with each other especially in the core. 
This is the reason that 
bulk field-aligned longitudinal plasma motion (propagation of brightness)
is evident in the flux tube (cf., Fig.~5). The plasma 
at the locations 'A', 'B' of sausage-brightened knots moves with speeds of 113 km s$^{-1}$, and
125 km s$^{-1}$. While, the curvature on 'D' first appeared and thereafter shrinks and slides 
in northward direction along the flux tube with speed of 170 km s$^{-1}$. Therefore, this may be the indication of the 
dominant longitudinal field and aligned plasma motions in the  flux tube system locally, which  vanishes further the 
creation of static magnetic islands and confinement of the plasma under the regime of 
sausage-pinch instability.

Some other alternative mechanisms may also play role in the formation of the 
observed bright plasma knots. 
Intially, the kinked flux tube 
may undergo in an internal reconnection and energy release. 
this 
The bursty reconnection due to the initial kink instability may also cause the 
formation of bright knots that could be the $"$magnetic islands$"$ 
formed by internal reconnection \citep{Kliem2000}. The formation 
of such magnetic islands in resistive corona is reported to occur 
quasi-periodically at comparatively larger spatio-temporal scales under the scenario 
of recurrent and dynamic reconnection in the current sheet \citep{Kliem2000}. Such mechanism,
may play some role, however, the direct comparison is not possible in the present 
observational base-line as 
it deals the short spatio-temporal dynamics and 
do not have a signature of periodic occurrance of reconnection and the associted 
formation of the magnetic islands due to tearing mode instabilities. 
The balooning instability can also create plasma magnetic islands 
near the loop apex due to the difference in plasma beta at its upper
boundary as well as below maintained layers \citep{Shib99,Tsap2008}. 
Our observed morphological scenario of the formation 
of multiple bright knots (ripples) over the whole body of the 
flux tube for $<$60 s is entirely different from the evolution of the 
balooning instability as observed recently
\citep{Ku12}.

In conclusion, our observations provide the evidence of morphological 
evolution of sausage-pinch instability in the corona. Further observations are required to 
fully understand its behaviour in corona. 
These observations may have implications to forthcoming theoretical and 
observational studies in understanding the role of such rare instability 
processes. More analysis should be performed in this 
direction using future high-resolution ground- and space based observations, 
which will provide new clues to the existing theories of sausage-pinch 
instability in magnetic flux tubes. 

\section{Acknowledgments}
We thank referee for the constructive suggestions received that improved the 
paper.
We acknowledge the use of the SDO/AIA observations for this study. 
RE acknowledges M. K\'eray for patient encouragement
and is also grateful to NSF, Hungary (OTKA, Ref. No. K83133) for financial support
received. 
AKS also thanks Shobhna Srivastava for her support and encouragement.
VF thanks Dr D. Vasylyev for his help with images preparation.
DT acknowledge the support from DST under Fast Track Scheme (SERB/F/3369/2012-2013). AKS highly acknowledges 
the adoptation of the method of E. Verwichte in making the Time-Distance diagrams along the curved paths.

{}

\clearpage

%
\clearpage

\clearpage

\begin{figure*}
\centering
\mbox{
\includegraphics[scale=0.25]{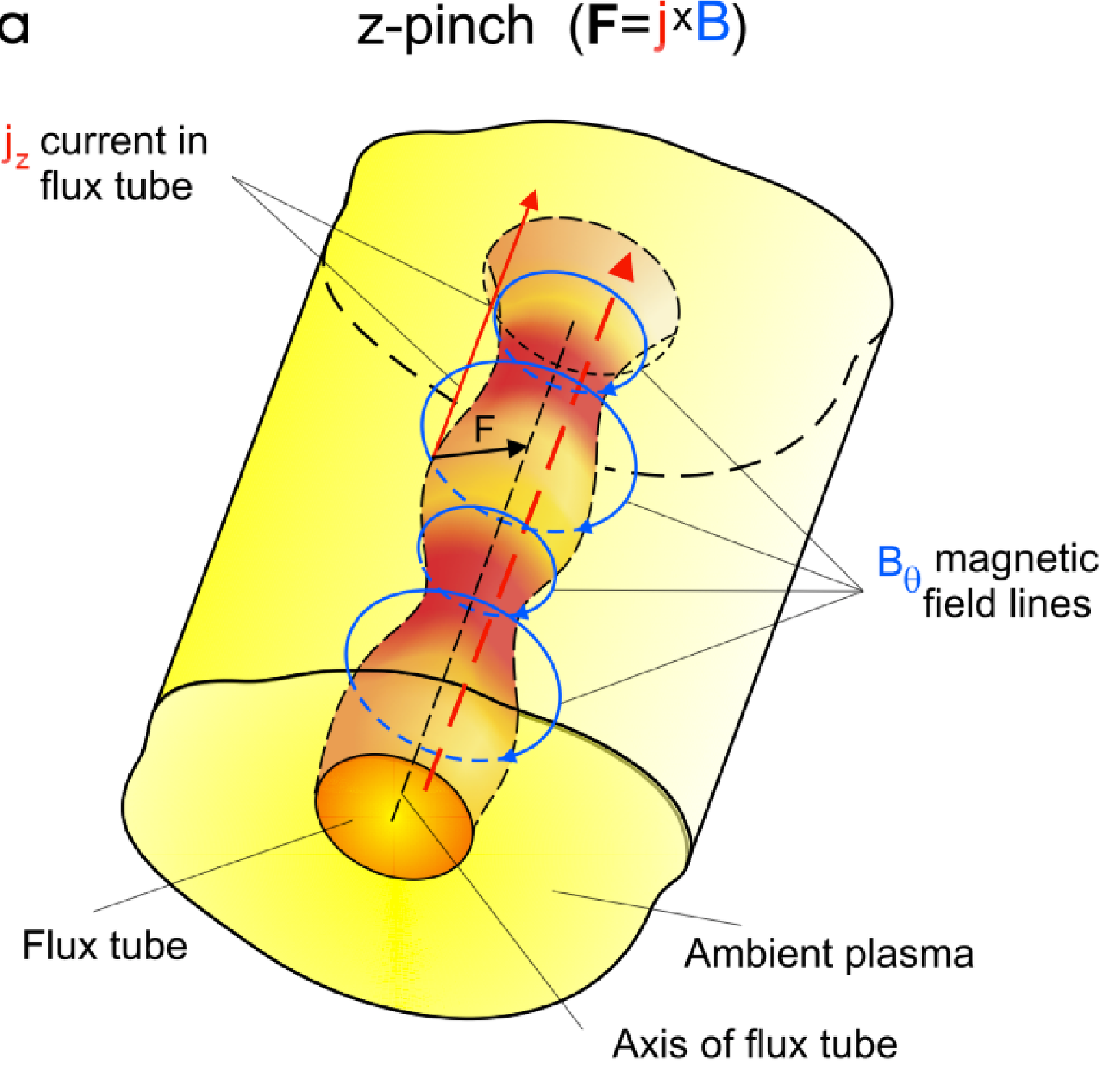}
\includegraphics[scale=0.25]{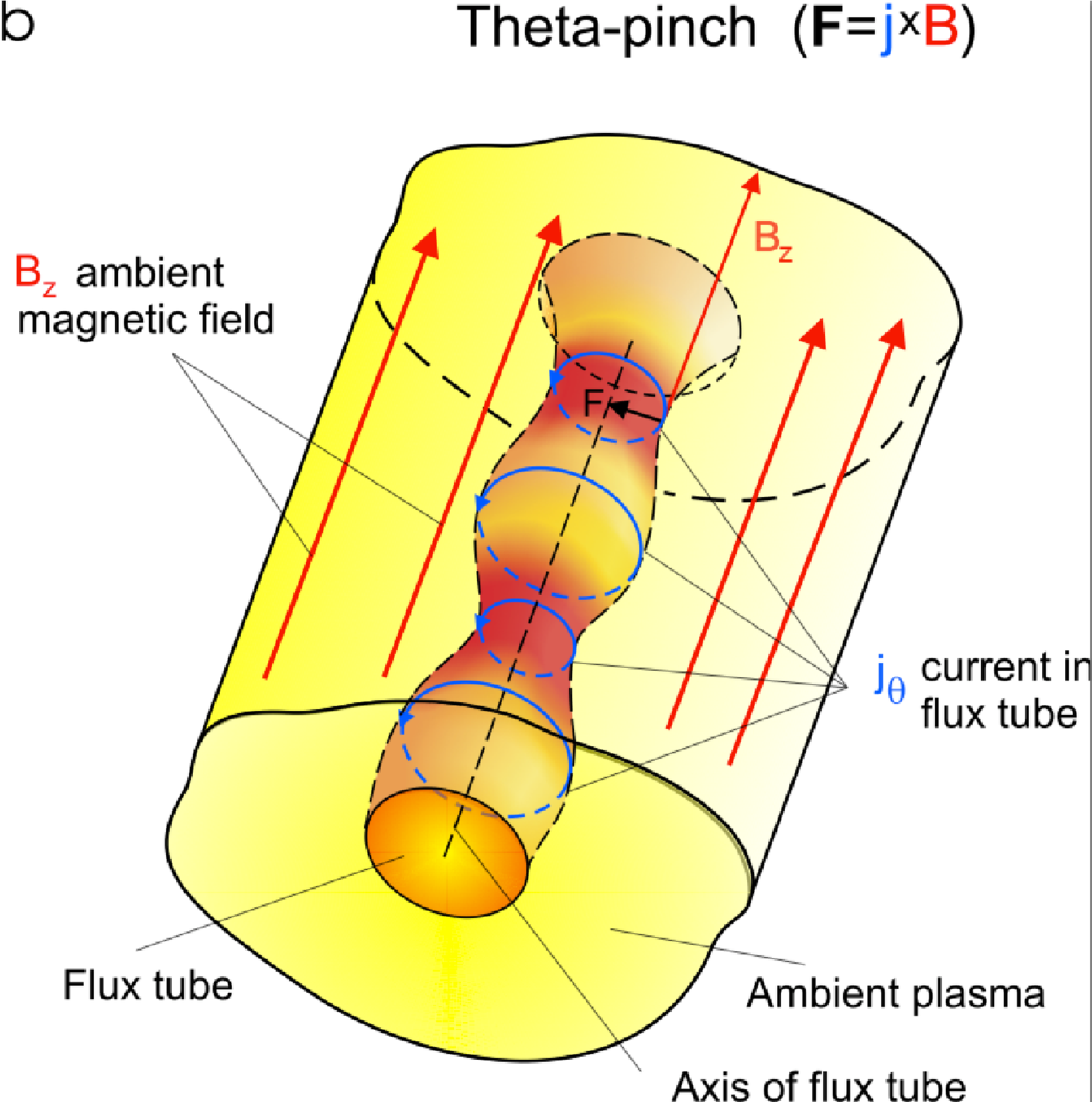}}
\mbox{
\includegraphics[scale=0.43]{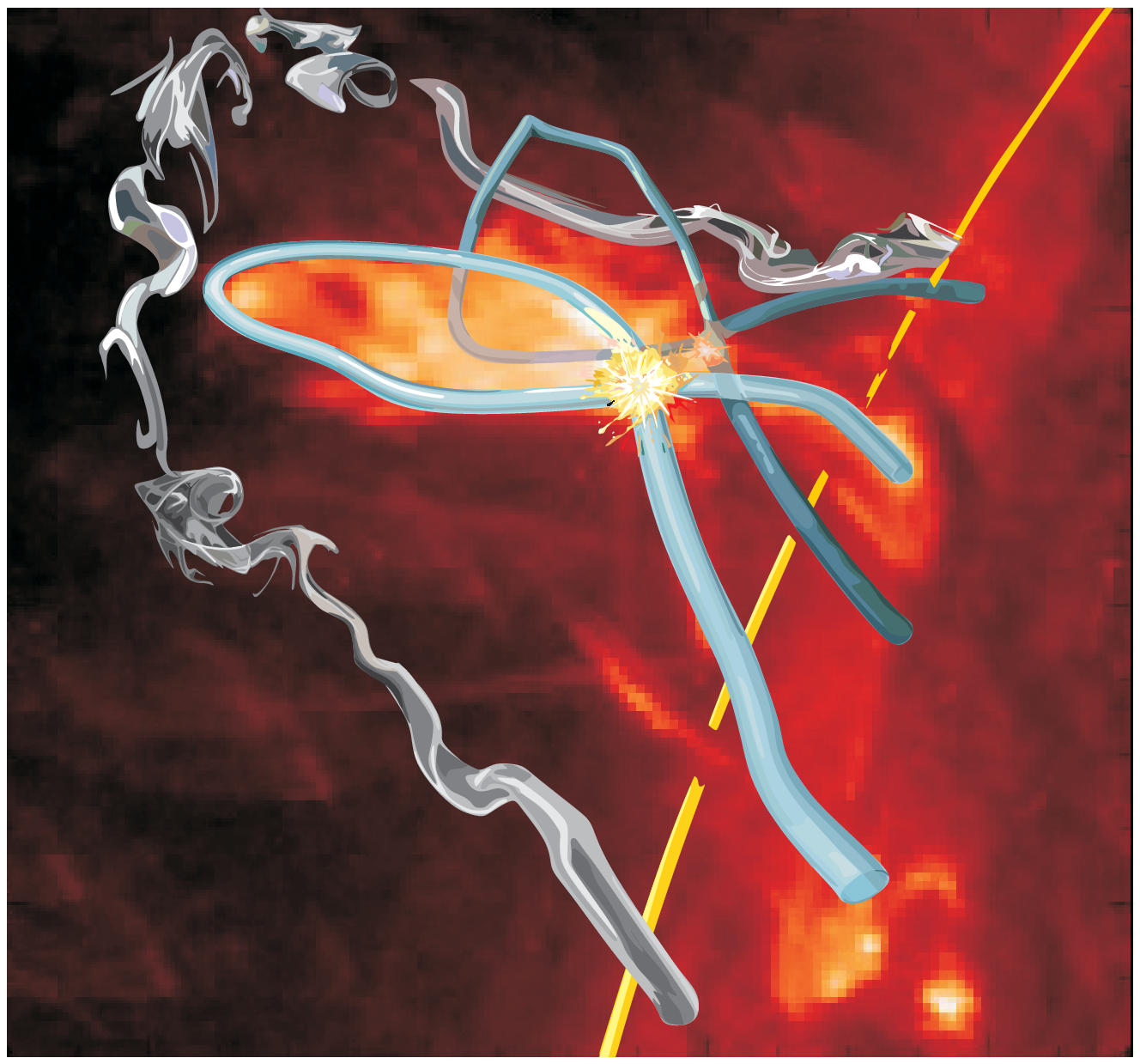}
\includegraphics[scale=0.43]{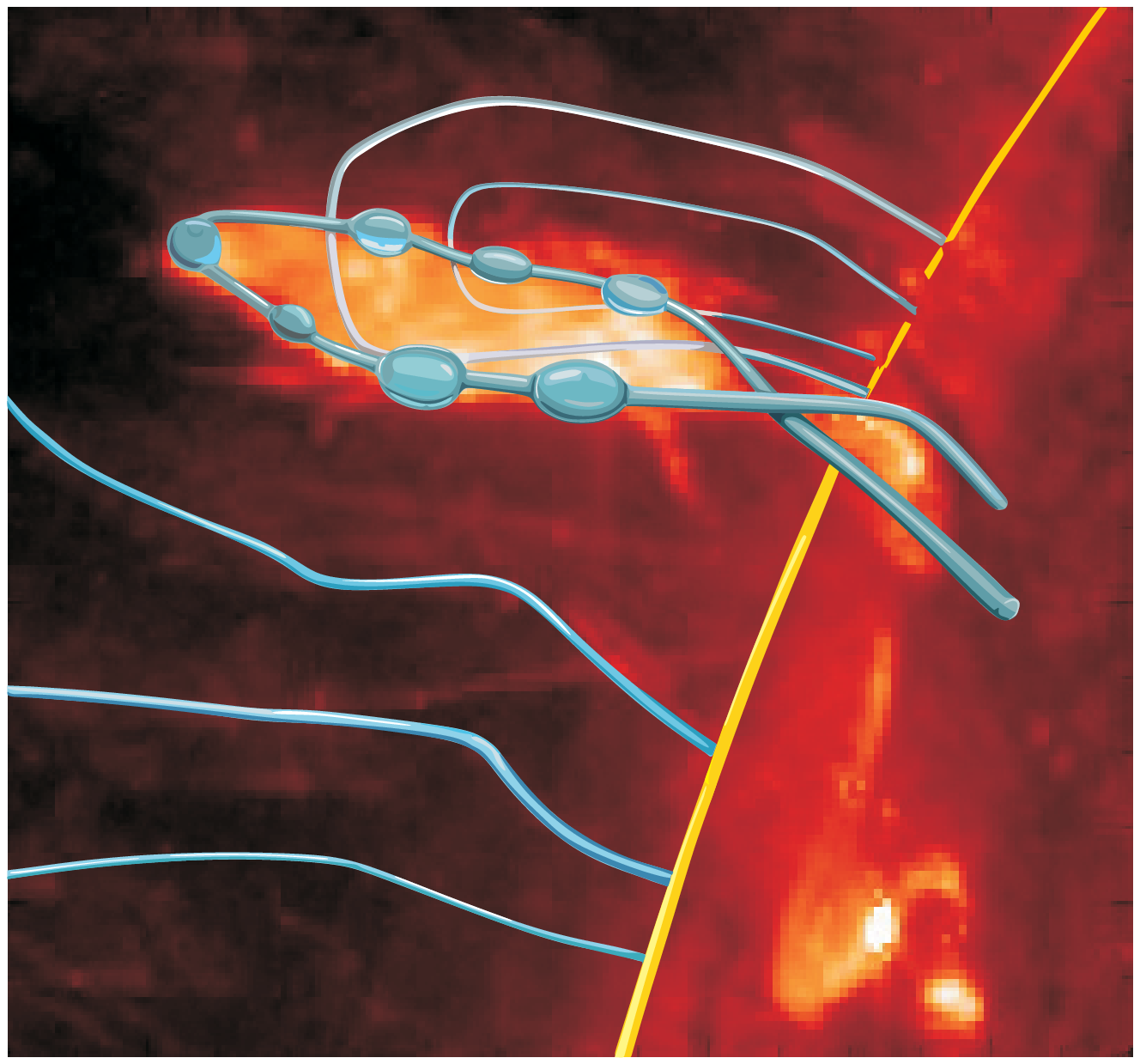}
\includegraphics[scale=0.43]{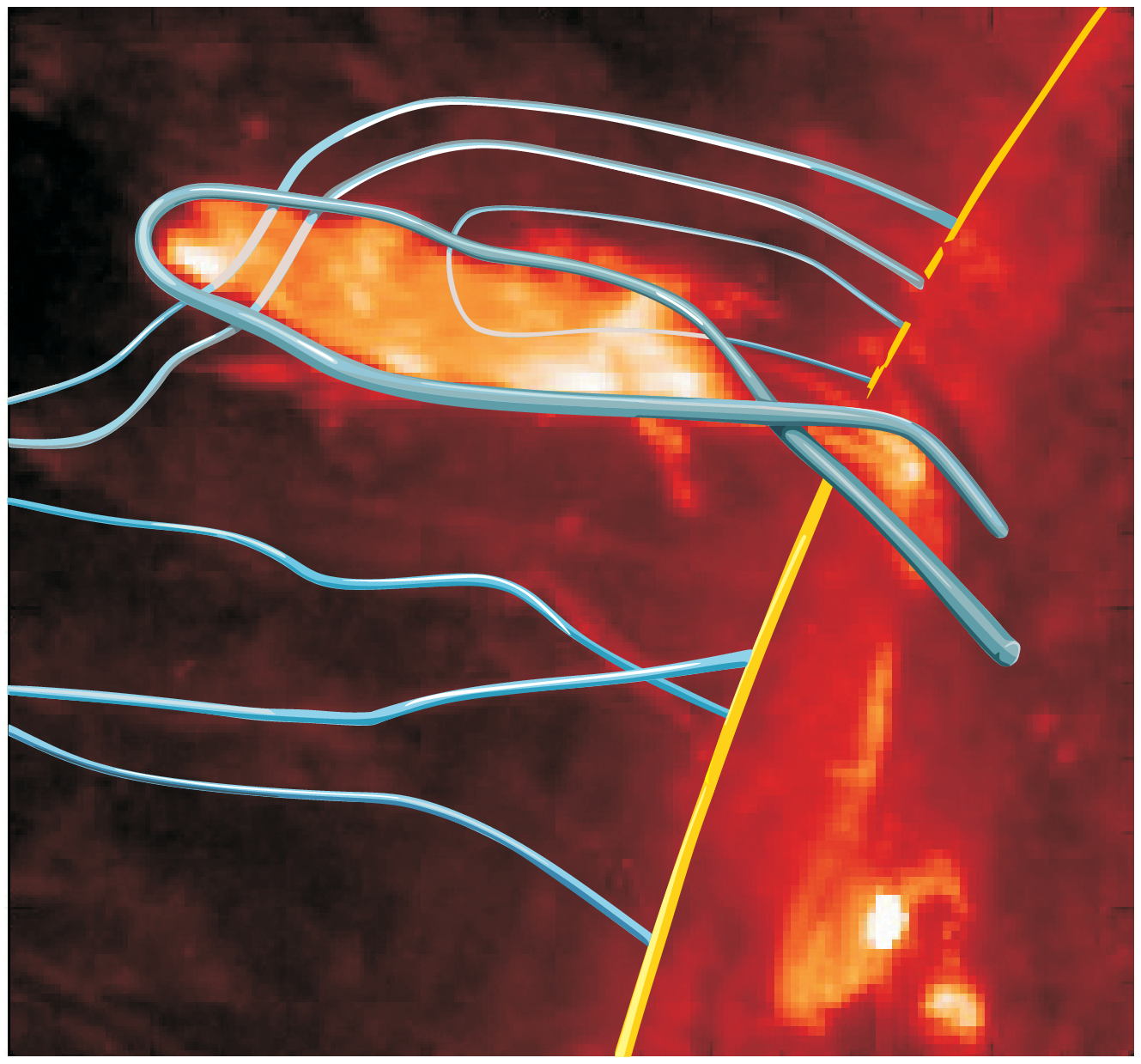}
}
\caption{\small Top-panel : Sketch of the physical mechanism of sausage 
instability in {\it z}-pinch (a) and $\theta$-pinch (b). 
In the case (a), when an $m=0$, i.e. sausage perturbation is 
superimposed on the flux tube, the magnetic field, 
$B_\theta$, (shown as blue cylindrical arrows) 
in the neck region grows due to the current 
$j_{z}$ (shown as dashed red arrow inside the magnetic flux tube) 
increasing in a smaller cross section. 
Additional magnetic pressure produces a force ({\bf F}) 
which tends to constrict the plasma cylinder. 
Bottom-panel : The handmade sketches overlaid on AIA 304 \AA\  snapshots
to explain the observed scenario. The observed localized 
dynamics of the fluxtube is generated by either of these
two physical mechanisms, i.e., $\theta$ or z-pinch.}
\label{fig:SKETCH}
\end{figure*}
%
\clearpage

\begin{figure*}
\centering
\mbox{
\includegraphics[height=6.0cm]{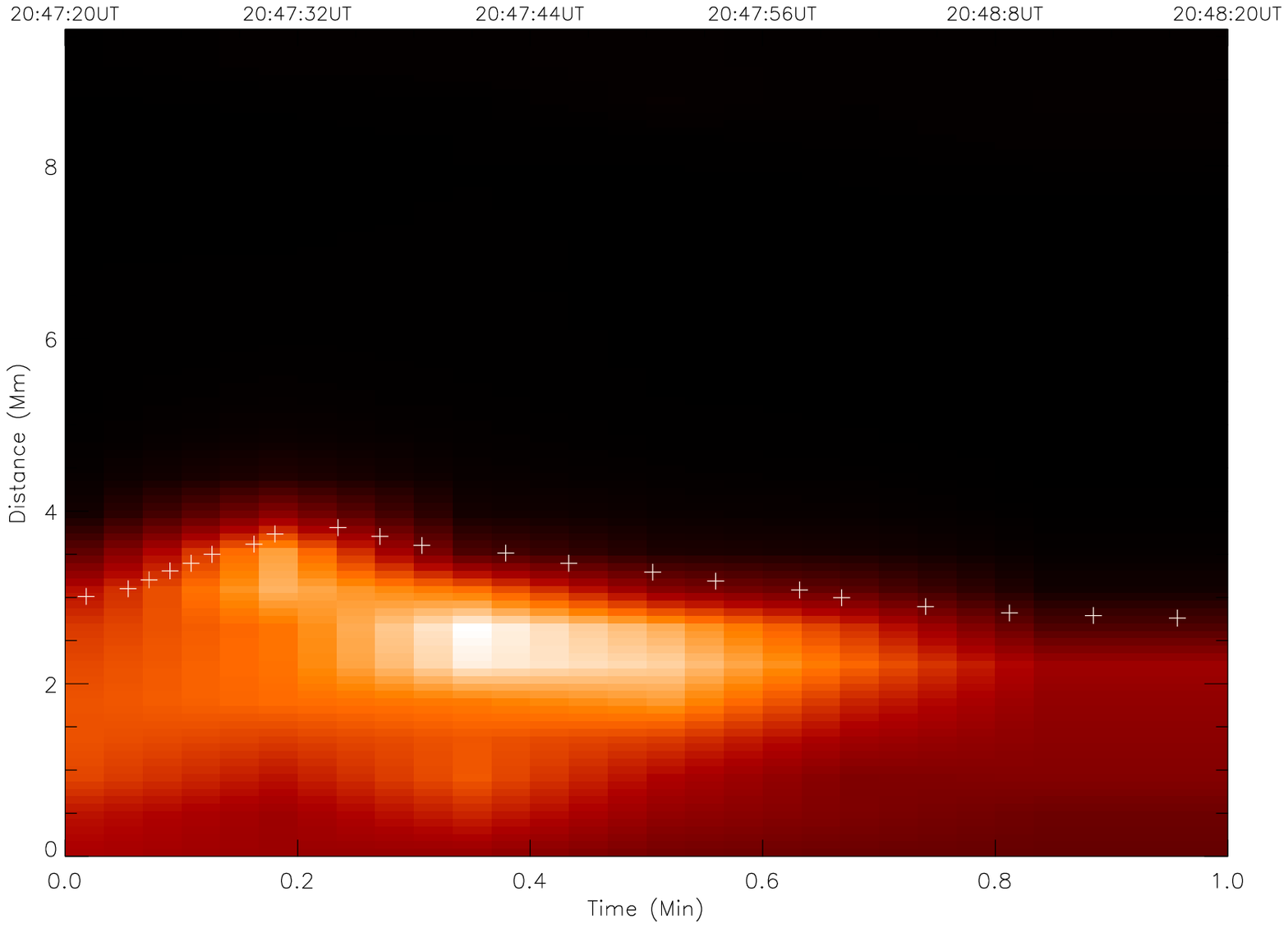}\llap{\raisebox{2.5cm}
{\includegraphics[height=2.5cm]{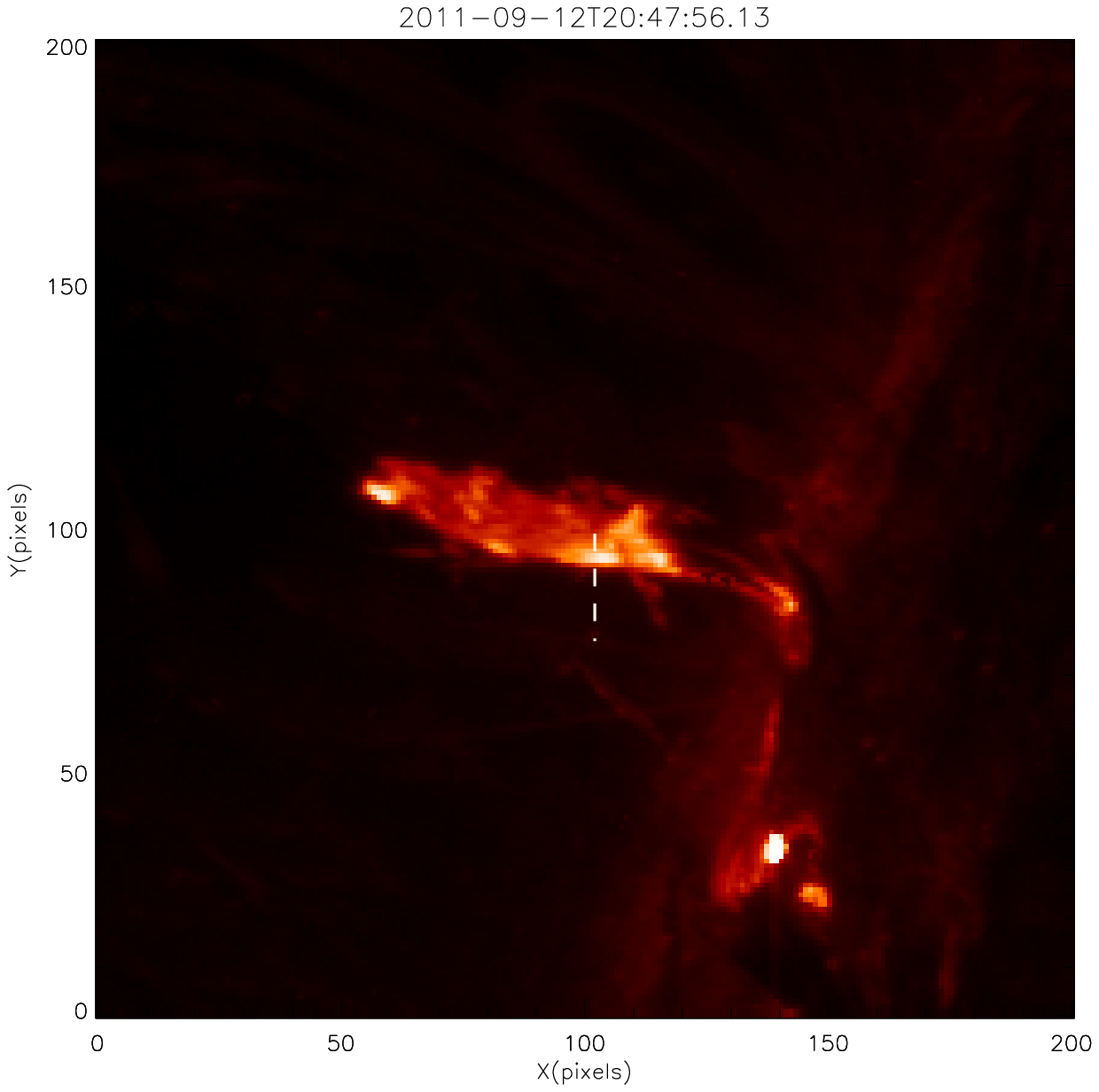}}}
\includegraphics[height=6.0cm]{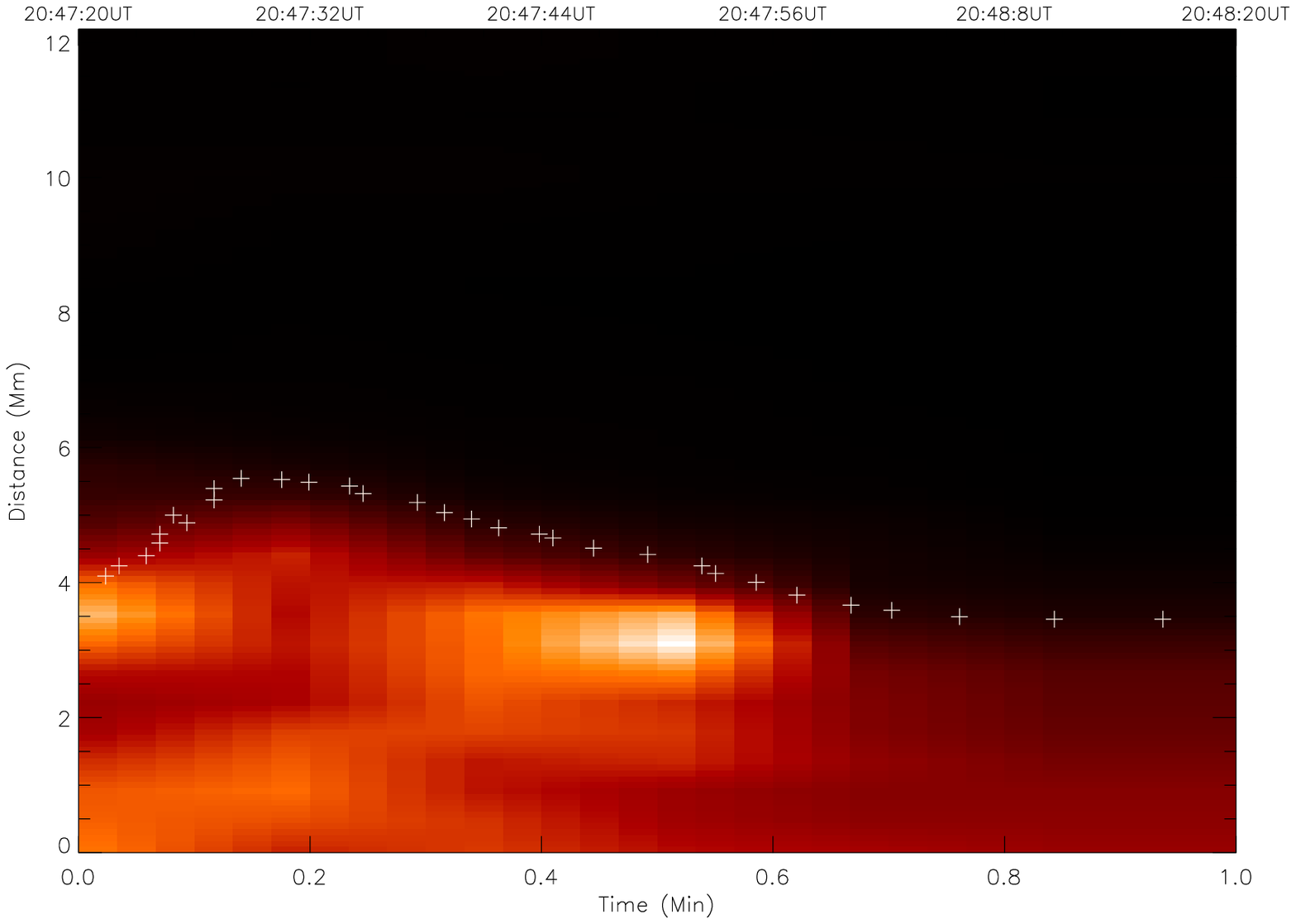}\llap{\raisebox{2.5cm}
{\includegraphics[height=2.5cm]{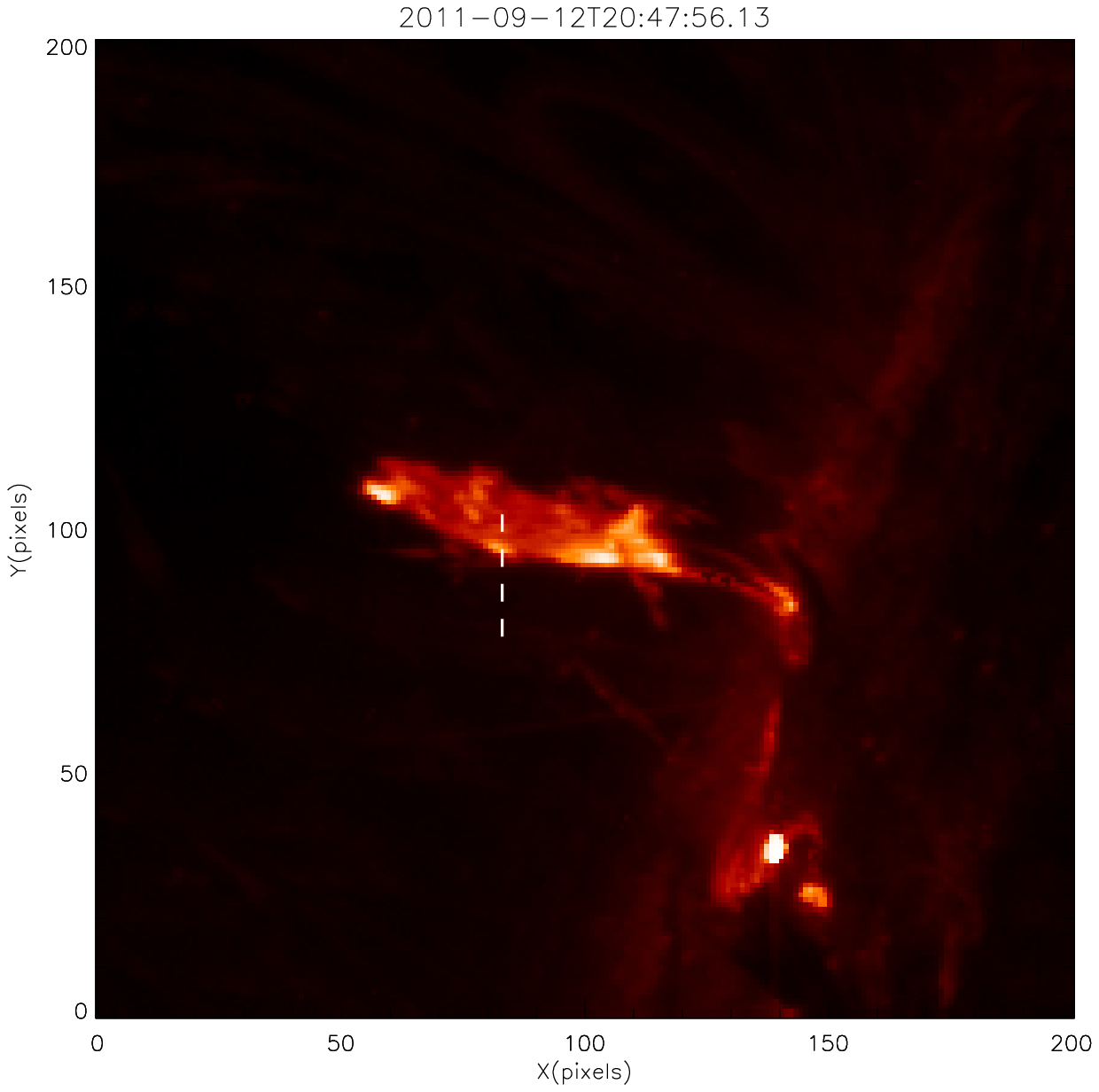}}}
}
\mbox{
\includegraphics[height=6.0cm]{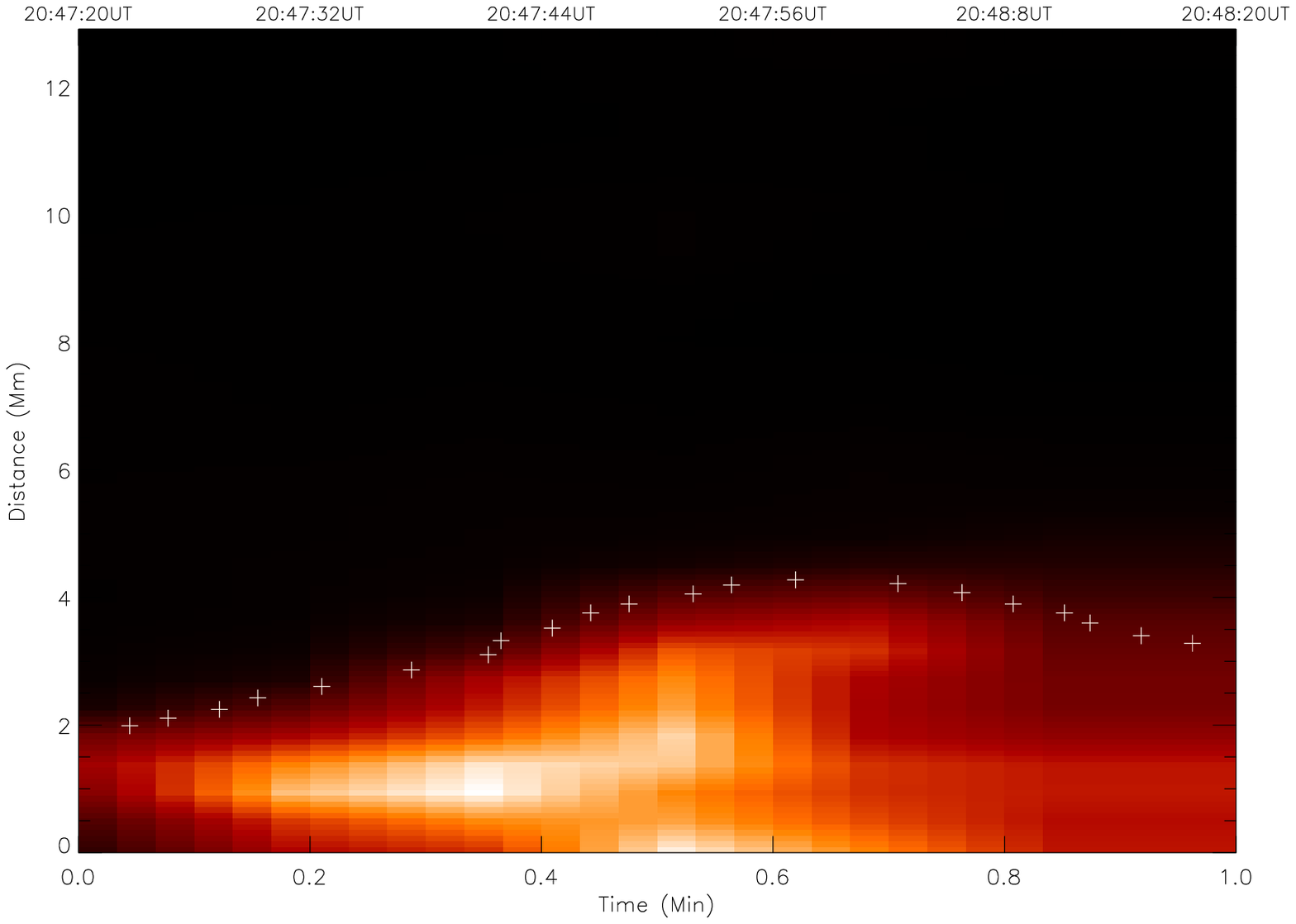}\llap{\raisebox{2.5cm}
{\includegraphics[height=2.5cm]{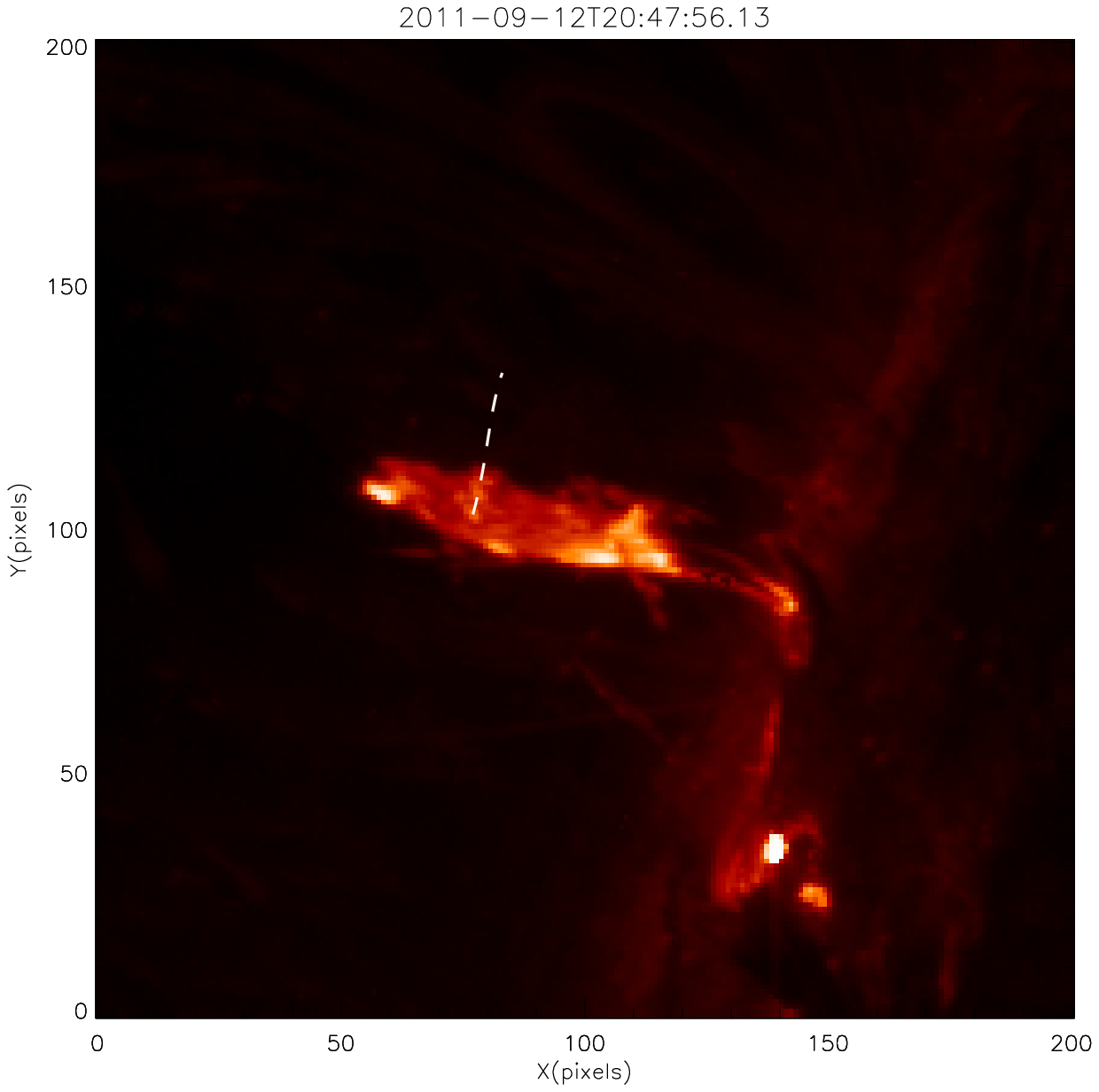}}}
\includegraphics[height=6.0cm]{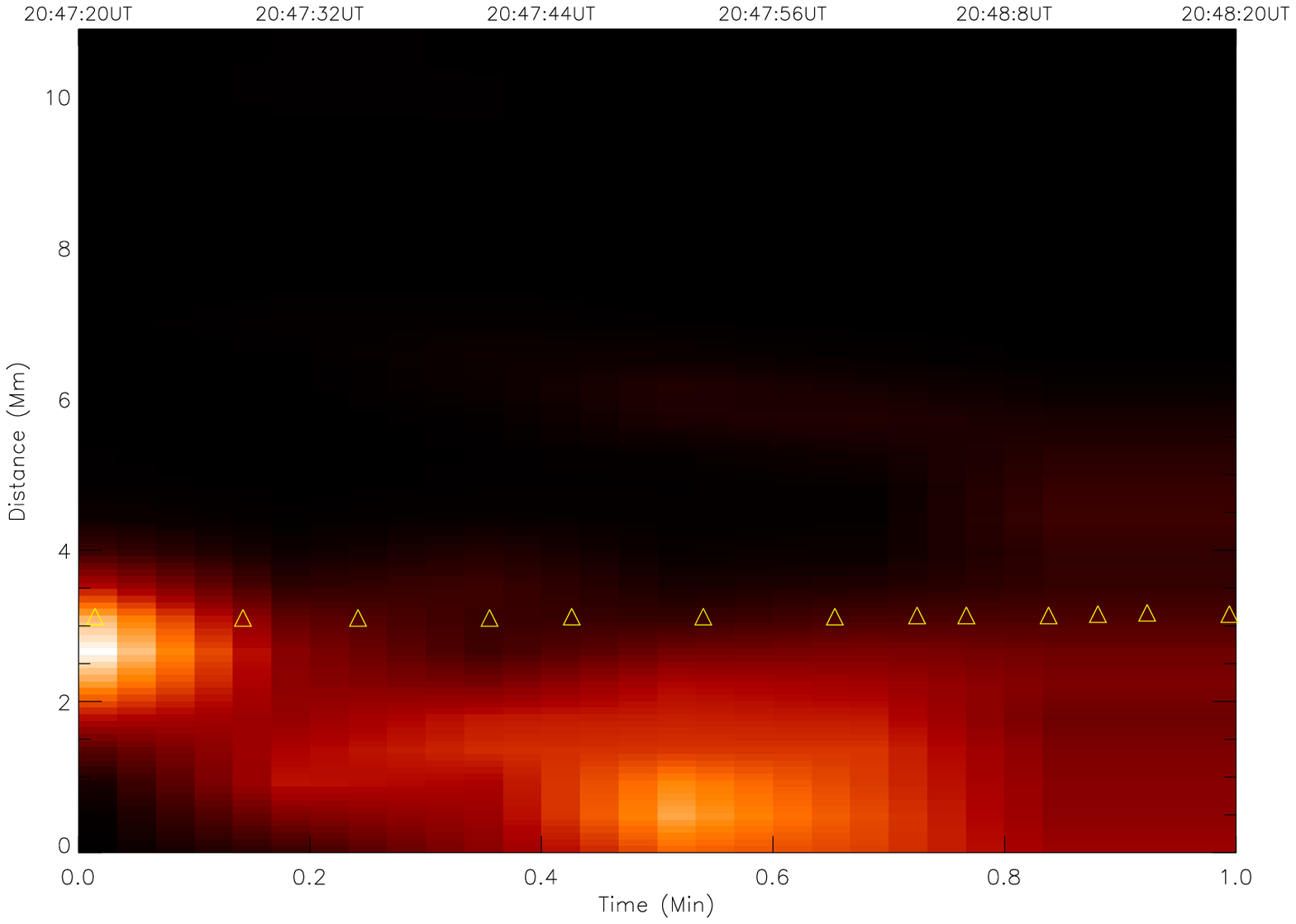}\llap{\raisebox{2.5cm}
{\includegraphics[height=2.5cm]{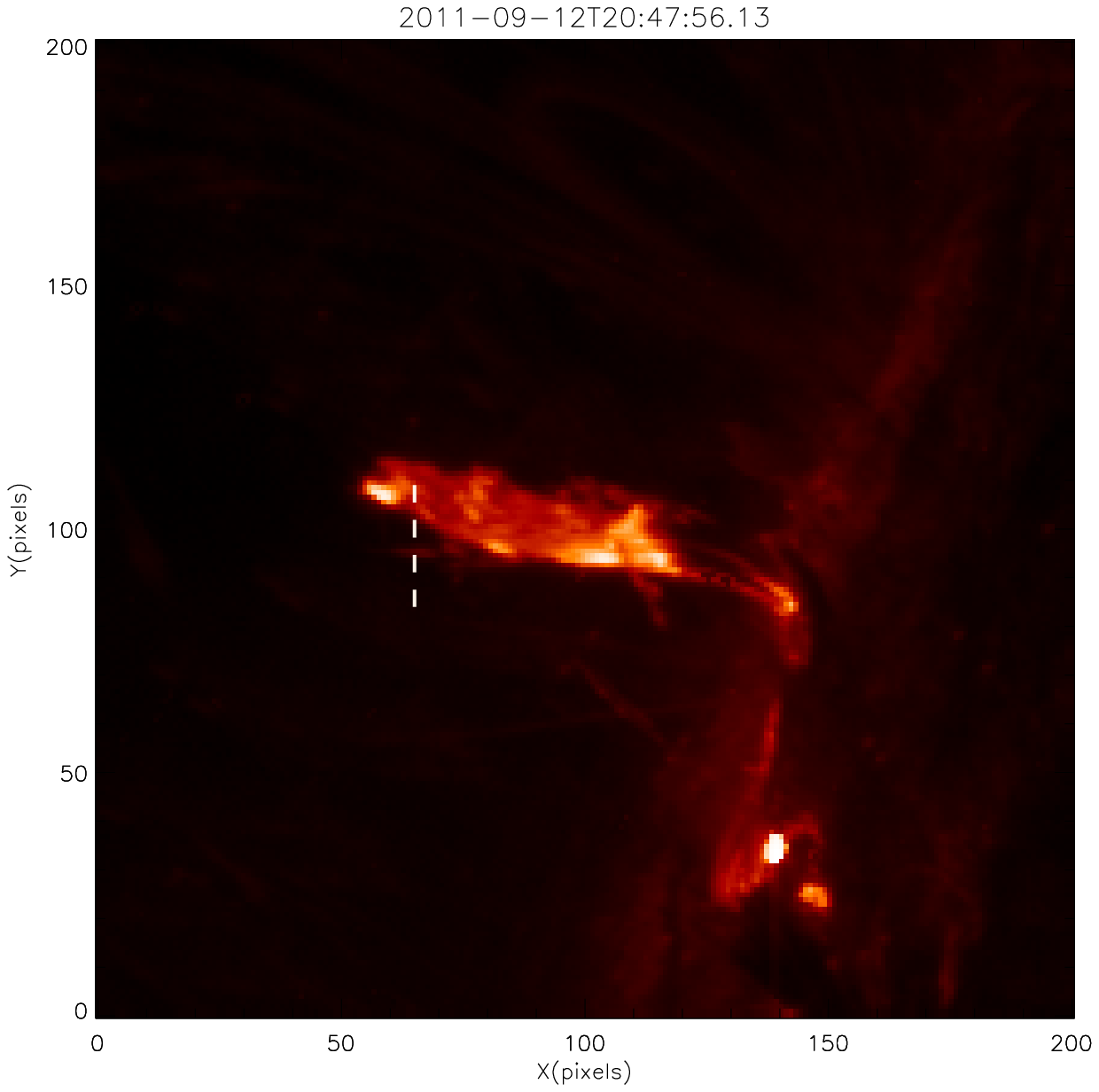}}}
}
\caption{\small
Distance-Time diagrams in AIA 304 \AA\ along the slits drawn across the various
surface curvatures over the flux tube marked by position 'A', 'B' (on southward part), 
'D' (on northward part). Last panel shows almost none variation of the surface of clearly evolved narrow neck between knot B and C
(Fig.~2).
}
\label{fig:JET-PULSE_3}
\end{figure*}
%
\clearpage

\begin{figure*}
\centering
\mbox{
\includegraphics[height=6.5cm]{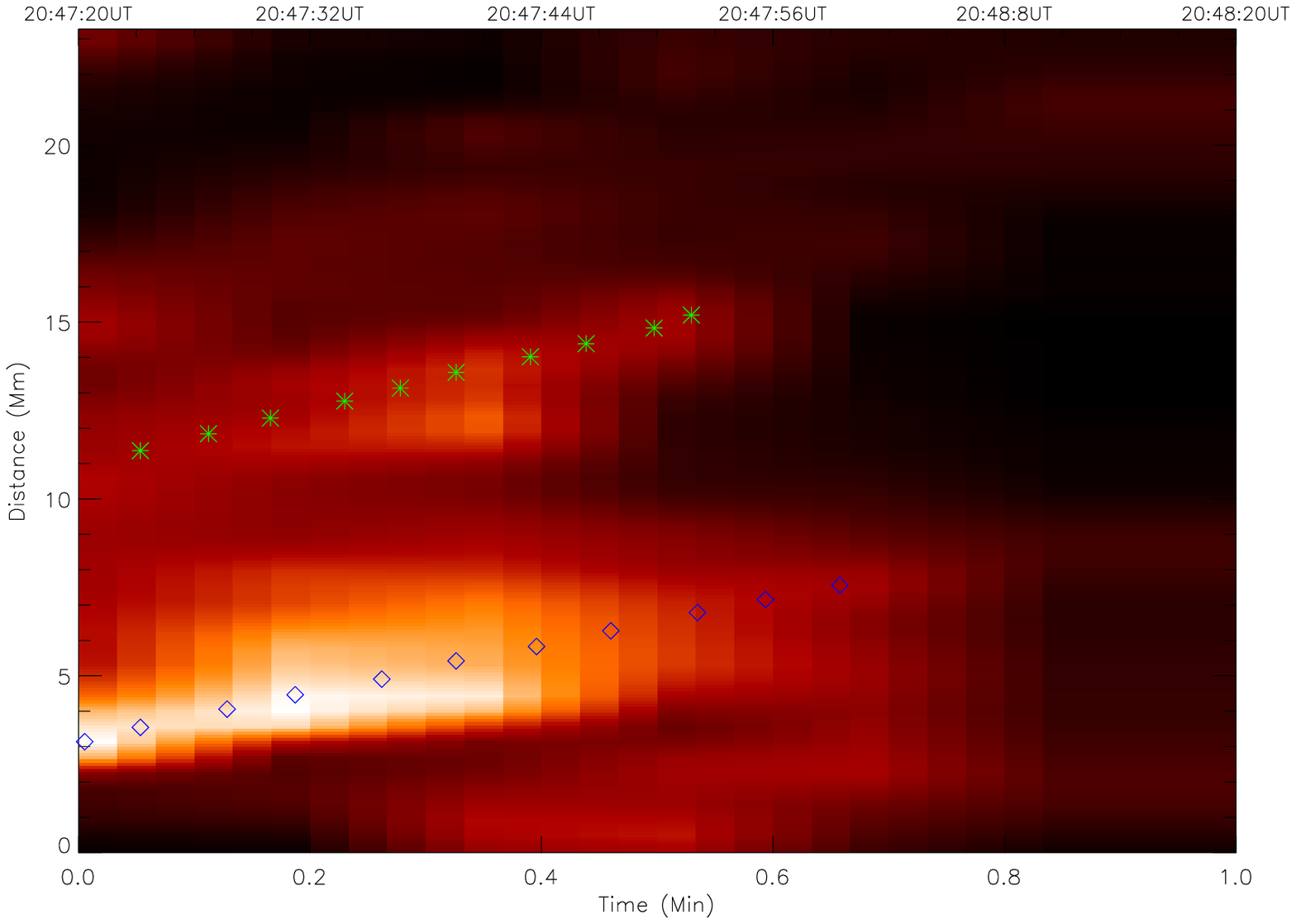}\llap{\raisebox{2.5cm}
{\includegraphics[height=2.5cm]{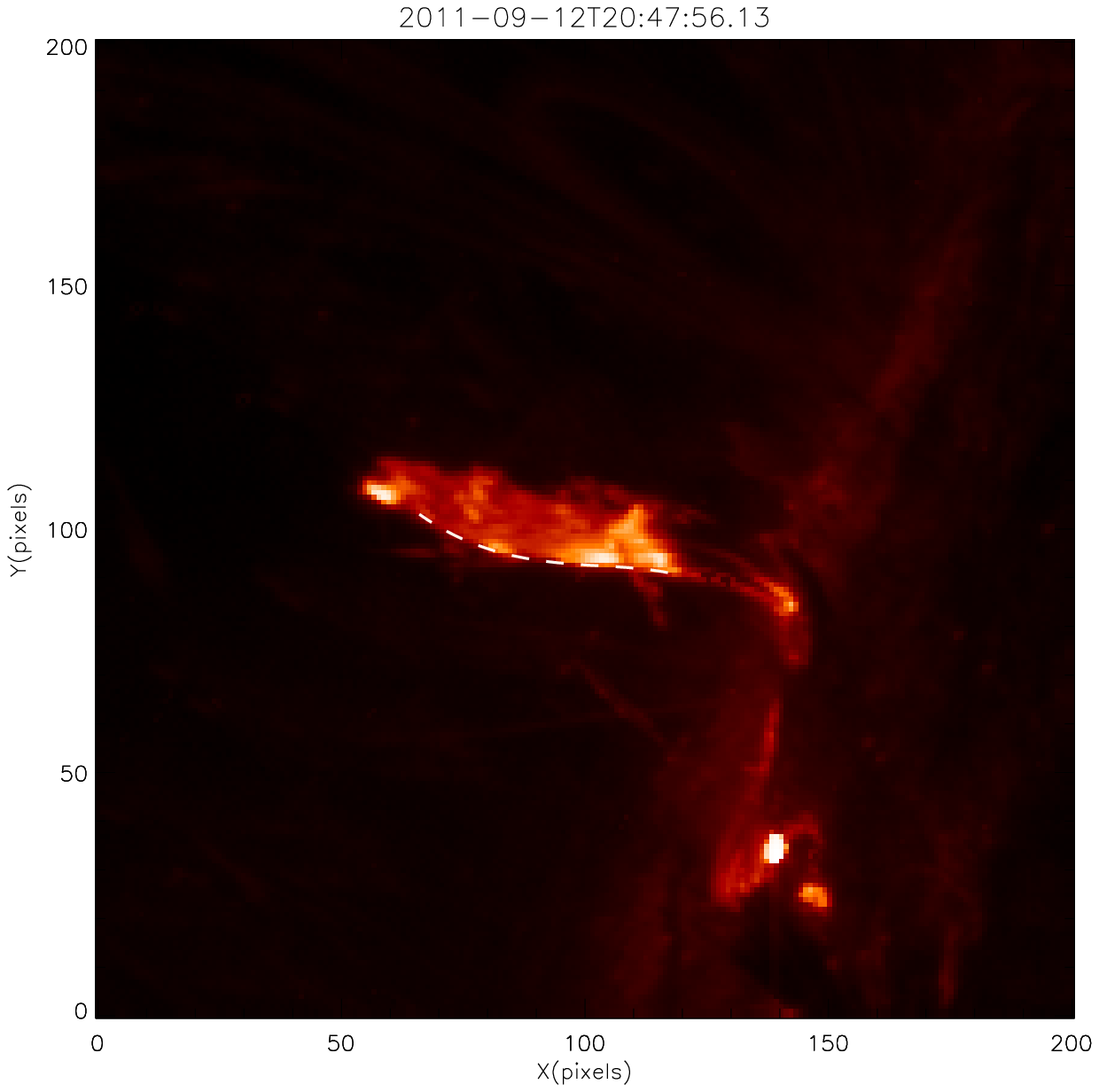}}}
\includegraphics[height=6.5cm]{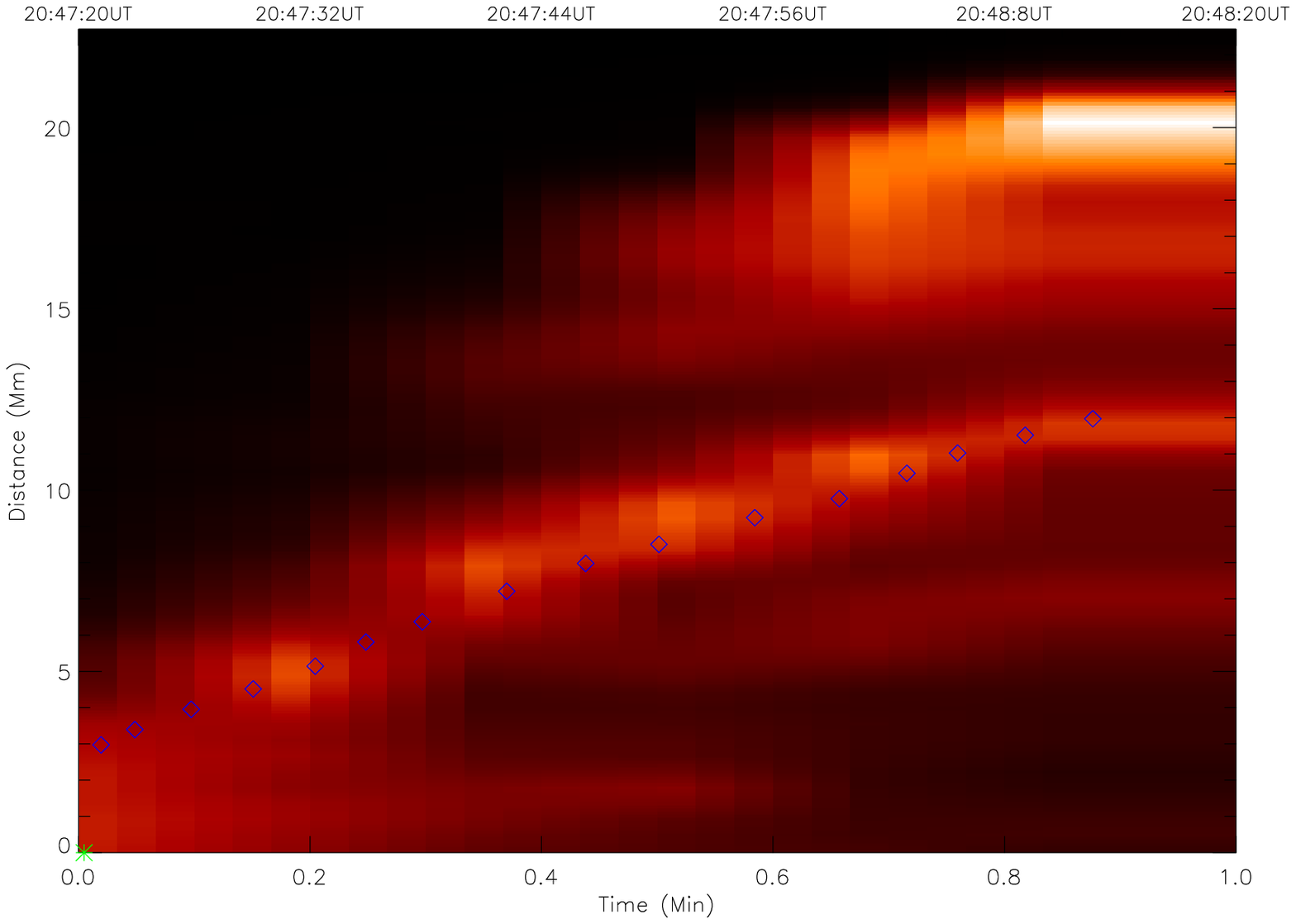}\llap{\raisebox{2.5cm}
{\includegraphics[height=2.5cm]{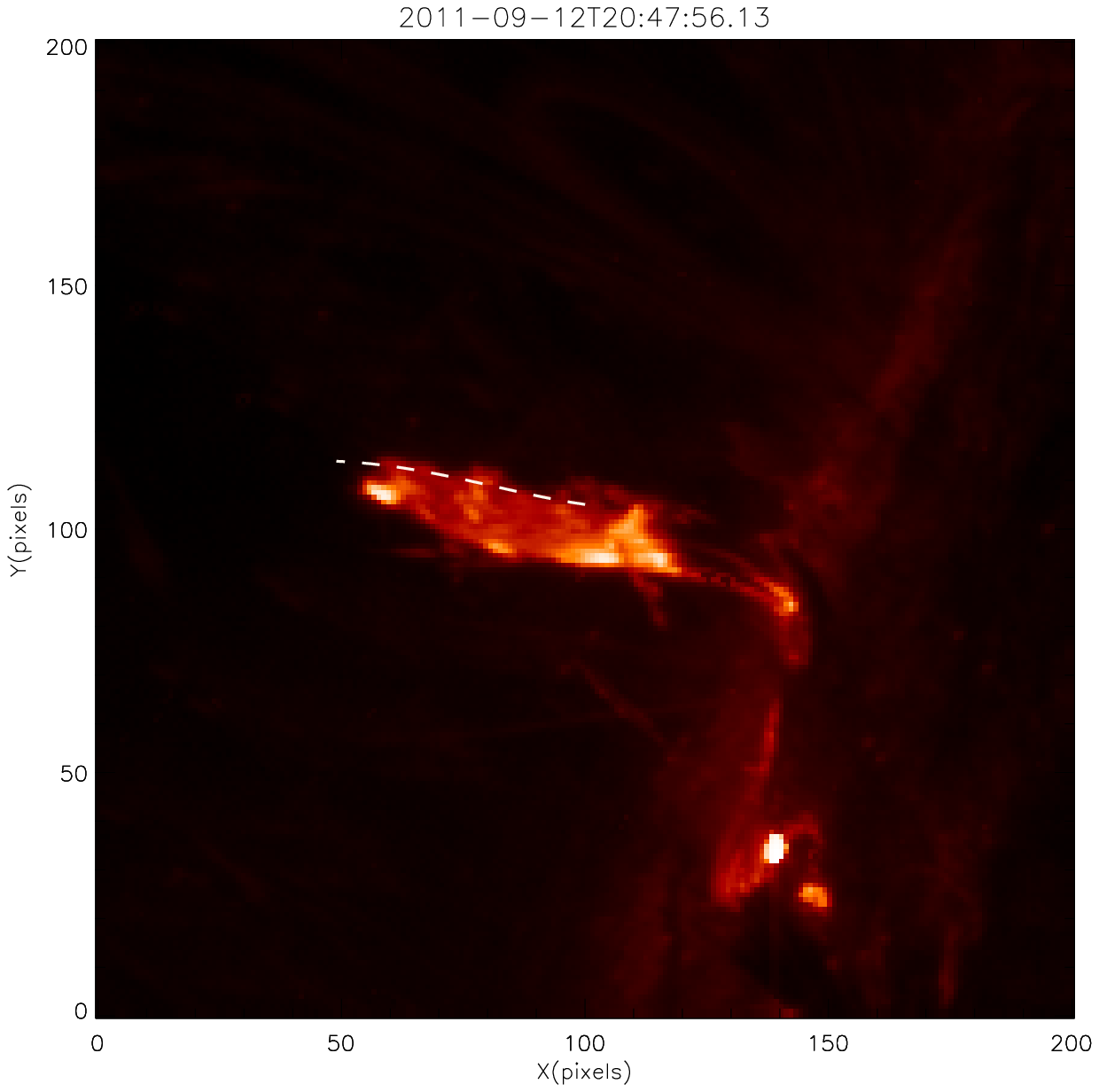}}}
}
\caption{\small
Distance-Time diagram in AIA 304 \AA\ along the curved path drawn along the southward 
part of the enveloping flux tube that shows the plasma (brightness) motion through two 
knots as marked by position 'A' (diamonds) and 'B' (stars), as well as on its northward part that consists 
the dynamic and bright knot 
'D' (cf., motion of brightness as marked by diamond symbols) in Fig.~2.
}
\label{fig:JET-PULSE_3}
\end{figure*}

\clearpage

\end{document}